\documentclass[showpacs,aps,prb,reprint,twocolumn]{revtex4-1}

\usepackage{commath}
\usepackage{amsmath}
\usepackage{graphicx}
\usepackage{hyperref}
\usepackage{txfonts}

\draft 

\begin{document}


\title{A universal macroscopic theory of surface plasma waves and their losses} 


\author{Hai-Yao Deng}
\email{haiyao.deng@gmail.com}
\thanks{Currently at National Graphene Institute, University of Manchester, UK.}
\affiliation{School of Physics, University of Exeter, EX4 4QL Exeter, United Kingdom}

\begin{abstract} 
Recently, we have revealed an intrinsic instability of metals due to surface plasma waves (SPWs) and raised the prospect of using it to create lossless SPWs. The counter-intuitive nature of this finding prompts one to ask, why had not this instability been disclosed before, given the long history of this subject? If this instability does exist, how far is it from reality? The present work is devoted to answering these questions. To this end, we derive a unified macroscopic theory of SPWs that applies to any type of electron dynamics, be they local or non-local, classical or quantum-mechanical. In light of this theory, we analyze the behaviors of SPWs according to several electron dynamics models, including the widely used local dielectric model (DM), the hydrodynamic model (HDM) and the specular reflection model (SRM), in addition to the less common semi-classical model (SCM). We find that, in order to unveil the instability, one must (i) self-consistently treat surface effects without any of the usually imposed auxiliary conditions and (ii) include translation symmetry breaking effects in electron dynamics. As far as we are concerned, none existing work had fulfilled both (i) and (ii). To assess the possibility of realizing the instability, we analyze two very important factors: the dielectric interfacing the metal and inter-band transitions, which both were ignored in our recent work. Whereas inter-band absorption -- together with Landau damping -- is shown adverse to the instability, a dielectric brings it closer to occurrence. One may even attain it in common plasmonic materials such as silver under not so tough conditions. 
\end{abstract}


\maketitle 

\section{Introduction}
\label{sec:1}
Electron density ripples propagating along metal surfaces, known as surface plasma waves (SPWs), have been intensively pursued as a promising enabler of nano photonics~\cite{barns2003,zayats2005,maier2007,sarid2010}. A fundamental issue hampering further progress is concerned with energy losses~\cite{khurgin}. SPWs lose energy fast due to Joule heating, Landau damping and inter-band absorption alongside others such as radiation losses~\cite{khurgin}. It has been suggested that these losses -- especially those due to Landau damping -- are intrinsic and cannot be significantly reduced without the addition of external gain medium~\cite{khurgin,oulton,stockman2008,hess2012,premaratne2017}. They would ultimately handicap the functionalities of SPWs~\cite{khurgin,oulton}. 

Our recent work~\cite{deng2017a,deng2017b,deng2017c} challenged the above view and found that, thanks to an incipient instability, the losses may well be reduced to any level without taking energy from outside. A critical point was shown to exist at $\gamma_0\tau = 1$, where $\tau^{-1}$ is the thermal electronic collision rate and $\gamma_0$ is a positive-definite quantity. At the critical point, the energy released from the Fermi sea -- at rate $\gamma_0$ -- just compensates for the dissipation due to electronic collisions and SPWs become lossless~\cite{deng2017c}. This instability was revealed through a semi-classical model (SCM, based on Boltzmann's transport equation) of SPWs. Given the highly counter-intuitive nature of this possibility and the long history on this subject dated back to late 1950s~\cite{ritchie1957}, the following questions must then be answered: \textit{Why has the criticality not been discovered in earlier work and what is missing in those work?} While it is not widely used in the study of SPWs, the SCM had certainly been considered before~\cite{harris1972,garcia1977b}. Yet nobody had claimed lossless SPWs.   

The main purpose of the present paper is to answer the above question. Our strategy is to construct a universal theory of SPWs that is applicable to any type of electron dynamics, be they local or non-local, classical or quantum-mechanical. In light of this theory, the properties of SPWs are then analyzed with several common electron dynamics models, including the local dielectric model (DM)~\cite{ritchie1957}, the classical hydrodynamic model (HDM)~\cite{ritchie1963,harris1971,barton1979,fetter1986} and the specular reflection model (SRM)~\cite{ritchie1966,ritchie1980,vagov2016} in addition to the SCM. The chief result obtained through this analysis is that, in order to unveil the instability, (i) a SPW theory must self-consistently deal with physical surfaces, and (ii) translation symmetry breaking effects must be included in the electron dynamics. Amongst these models, only the SCM is capable of (ii), but previous work based on the SCM had all failed with respect to (i). This explains why the existing work had failed to hit upon the possibility of lossless SPWs. 

In existing work based on non-local models, auxiliary conditions have usually been imposed~\cite{pekar1957,maradudin1973,garcia1977a,garcia1979}, mostly assuming no normal current on surfaces, as in the hard-wall picture adopted in computational studies~\cite{feibelman1982,tsuei1991,apell1984,pitarke2001}. From a conceptual point of view, however, these conditions are obviously incompatible with local models and arise only due to an incomplete description of physical surfaces~\cite{ginzberg1966}. We will show that the conventional treatment of SPWs via auxiliary conditions does not recover the standard textbook SPWs in the local models. We prescribe a simple yet complete macroscopic description of physical surfaces, which remedies the conceptual deficiency and allows to derive a universal SPW theory accounting for surface effects self-consistently.  

As a secondary purpose, we wish to address an experimentally interesting issue, that is, \textit{how far is the instability from reality and how can it be achieved?} In general, $\tau^{-1}$ is comparable to the characteristic plasma frequency $\omega_p$, even in defect-free materials and at zero temperature~\cite{beach1977}. This is because $\tau^{-1}$ is the collision rate at the SPW frequency, which is in the order of a few eVs in metals and thus effects as an effective high temperature of tens of thousands of Kelvins opening up a large phase volume for electron scattering. One then expects $\gamma_0\tau < 1$ usually. In order to devise a practically useful method of enhancing $\gamma_0$, we employ our theory to analyze two important factors affecting $\gamma_0$: a dielectric interfacing the metal supporting SPWs and inter-band transition effects, which were ignored in Refs.~\cite{deng2017a,deng2017b,deng2017c}. We find that, inter-band absorption (as well as Landau damping) strongly reduces $\gamma_0$, whereas topping a dielectric can significantly increase $\gamma_0$. In this way the instability may well be in reach in a single crystal of silver.  

This paper is organized as follows. We develop a universal macroscopic SPW theory in the next section and apply it to the DM, the HDM and the SRM in Sec~\ref{sec:3}. Section~\ref{sec:2.5} includes a discussion of a phenomenological approach to inter-band transition effects to make the paper self-contained. Section \ref{sec:4} is devoted to a thorough treatment of SPWs within the SCM. Dielectric and inter-band effects are analyzed. We assess the possibility of creating lossless SPWs by the SCM and conclude the paper in Sec.~\ref{sec:5}. In Appendices~\ref{sec:a1} and \ref{sec:a2}, we discuss some historical misconceptions regarding the SPWs in the HDM and the SRM, respectively.
 
\begin{figure}
\begin{center}
\includegraphics[width=0.45\textwidth]{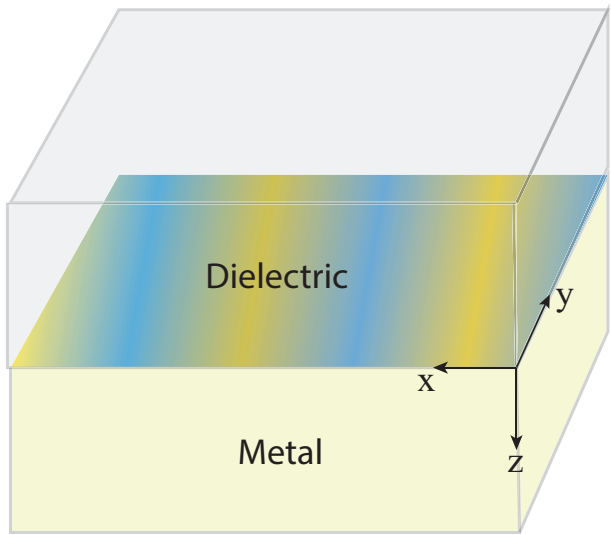}
\end{center}
\caption{Sketch of the system: SPWs are propagating along the interface between a metal and a dielectric. A point in space is denoted by its position vector $\mathbf{x} = (\mathbf{r},z)$, where $\mathbf{r} = (x,y)$ is the planar projection. The position of a point on the interface is denoted by $\mathbf{x}_0 = (\mathbf{r},0)$. \label{fig:f1}}
\end{figure}

\section{Universal theory of SPWs}
\label{sec:2}
In this section, we present a general theory of charge density waves in semi-infinite metals (SIMs). Extension to systems of other geometries is straightforward and will be considered elsewhere. The theory is formulated in terms of universal physical concepts and makes no resort to the particulars of electron dynamics. It is a macroscopic theory, thus valid as long as the thickness ($d_s$, typically a few lattice constants) of the microscopic surface layer, which forms between the vacuum and the bulk metal, is much smaller than the SPW wavelength $\Lambda$. In this paper, by 'macroscopic' we always mean $d_s/\Lambda \ll 1$, with no regard to electron dynamics. 

Retardation effects are neglected throughout this paper, which is reasonable provided the SPW phase velocity is much less than the speed of light in vacuum.

\subsection{Macroscopic description of surfaces}
\label{sec:2.1}
The SIM under consideration is assumed to occupy the region $z\geq0$ and bounded by a flat interface/surface macroscopically located at $z=0$. The other half space is either the vacuum or a dielectric with dielectric constant $\epsilon_d$, see Fig.~\ref{fig:f1}. We shall use the vector $\mathbf{x} = (\mathbf{r},z)$ to denote a point in space, where $\mathbf{r} = (x,y)$ is the planar projection of $\mathbf{x}$. Further, $\mathbf{x}_0 = (\mathbf{r},0)$ denotes a point on the surface and $t$ denotes time.

In a macroscopic description of SPWs, one usually considers the metal as a medium for an electromagnetic field and seeks surface localized (polariton) solutions of the governing Maxwell's equations~\cite{pitarke2007}. The procedure is to write down the waves for (an infinite metal) on the metal side and those (for an infinite dielectric) on the dielectric side, and then invoke conditions to match those waves at the surface. With local dynamics, the usual Maxwell's boundary conditions would do the job. With non-local dynamics, however, they are insufficient. Historically, auxiliary conditions -- mostly assuming zero normal current at the surface~\cite{harris1972,harris1971,garcia1977b,garcia1977a} -- have been invented as a remedy since 1950s~\cite{pekar1957}, which however treat the symptoms not the cause. The cause is a conceptual deficiency in the knowledge of surfaces~\cite{ginzberg1966}. Considering that a real microscopic surface can hardly be specified even for the simplest material due to preparation procedures, one might deem it hopeless to have a complete description. The following elementary analysis suggests otherwise.  

Let us imagine bringing two materials (A and B) in contact, and an interfacial layer of thickness $d_s$ -- in the order of a few lattice constants -- shall form in between. We may characterize this layer by a surface potential $\phi_s$, which should quickly decay to zero in the bulk regions outside the interfacial layer. The exact microscopic profile of the layer varies from one case to another and can hardly be known \textit{a priori}. Despite this, we may still write down a generic form for the electric current density $\mathbf{j}(\mathbf{x},t)$ in the whole system including the interfacial layer. To this end, we observe that in the bulk regions where $\phi_s$ vanishes, the form of $\mathbf{j}(\mathbf{x},t)$ can be completely specified with the respective dynamic equations for the infinite materials, apart from some parameters (such as the \textit{Fuchs} parameter, see Sec.~\ref{sec:4}) that encode the effects of surface scattering on the electron waves. Let us denote by $\mathbf{J}_{A/B}(\mathbf{x},t)$ the forms of $\mathbf{j}(\mathbf{x},t)$ in the bulk region of A/B. Microscopically, $\mathbf{j}$ evolves from $\mathbf{J}_A$ in the bulk region of A, through a rapid variation in the interfacial layer, to $\mathbf{J}_B$ in the bulk region of B. Formally, we can write for the $\mu$-th component of the current density that $j_\mu(\mathbf{x},t) = J_{A,\mu}(\mathbf{x},t) w_\mu(z) + J_{B,\mu}(\mathbf{x},t)(1-w_\alpha(z))$, where the profile functions $w_\alpha(z)$ approach unity in the bulk region of A and zero in that of B. The exact profile of $w_\mu(z)$ depends on the microscopic details of the interfacial layer. On the scale of $\lambda$, however, the interfacial layer appears infinitely thin and $w_\alpha(z)$ reduce to the Heaviside step function $\Theta(z)$, where $\Theta(z\geq0) = 1$ and $\Theta(z<0)=0$, in a macroscopic theory. One thus ends up with
\begin{equation}
\mathbf{j}(\mathbf{x},t) = \mathbf{J}_A(\mathbf{x},t)\Theta(z) + \mathbf{J}_B(\mathbf{x},t)(1-\Theta(z)). \label{sur}
\end{equation}
This holds valid for any $w_\mu(z)$ and is thus a general and complete macroscopic description of a physical interface, as long as a perturbation on one side does not cause significant responses on the other. 

To recapitulate, equation~(\ref{sur}) elegantly captures two importance physical consequences of an interface: the rapid variation of the current density through the step function $\Theta(z)$ and the surface scattering effects on electron dynamics through the parameters contained in the bulk forms. These scattering effects -- including the symmetry breaking effects -- have been ignored in most models except for the SCM. In general $\mathbf{J}_A$ and $\mathbf{J}_B$ are not equal on the interface, as is certainly the case for local dynamics models, and charges accumulate in the interfacial layer. Such capacitive effects would be mistakenly erased under auxiliary conditions, which often dictate continuity of current across the interface, e.g. the vanishing of normal current at the metal-vacuum interface.   

\subsection{Charge density waves}
\label{sec:2.2}
Now we derive the equations of motion for the charge density $\rho(\mathbf{x},t)$ in the SIM. Our starting point is the equation of continuity. Specifying Eq.~(\ref{sur}) to the SIM, we have $\mathbf{j}(\mathbf{x},t) = \Theta(z) \mathbf{J}(\mathbf{x},t)$, where $\mathbf{J}(\mathbf{x},t)$ is the current density in the metal. The equation of continuity then reads
$$\mathcal{D}_t\rho(\mathbf{x},t) + \partial_{\mathbf{x}}\cdot \mathbf{j}(\mathbf{x},t)= 0, \quad \mathcal{D}_t  = \tau^{-1}+\partial_t.$$ Here we have included a global relaxation term $-\rho(\mathbf{x},t)/\tau$ to account for the thermal relaxation of non-equilibrium charges due to microscopic electronic collisions driving the system toward thermodynamic equilibrium. In terms of $\mathbf{J}(\mathbf{x},t)$, this equation becomes
\begin{equation}
\mathcal{D}_t\rho(\mathbf{x},t)  + \partial_{\mathbf{x}}\cdot\mathbf{J}(\mathbf{x},t) = -\Theta'(z)J_z(\mathbf{x}_0,t), \label{2.1}
\end{equation}
where $\Theta'(z) = \partial_z\Theta(z)$. The right-hand term of this equation signifies the capacitive effects, which is critical in the energy conversion process but had been overlooked until our recent work~\cite{deng2017c}. This term was also noticed by A. L. Fetter in studying the edge plasmon in two-dimension systems~\cite{fetter1986}. 

Without loss of generality, we may assume a quasi-plane wave for the fields and write $\rho(\mathbf{x},t) = \rho(z)e^{i(\mathbf{k}\cdot{r}-\omega t)}$ and similarly for $\mathbf{J}(\mathbf{x},t)$ and the electric field $\mathbf{E}(\mathbf{x},t)$, where the wave vector $\mathbf{k} = (k\geq0,0)$ is directed along the positive $x$-axis for the sake of definiteness. In general the frequency $\omega = \omega_s - i\gamma$ is complex. The relation between $\omega$ and $k$ is determined by the wave equations to be established in what follows. Equation~(\ref{2.1}) thus becomes
\begin{equation}
-i\bar{\omega}\rho(z)+\nabla\cdot\mathbf{J}(z) = - J_z(0)\Theta'(z), \quad \nabla = (i\mathbf{k},\partial_z),\label{2.2}
\end{equation}
where $\bar{\omega} = \omega + i/\tau$ and $J_z(\mathbf{x}_0,t) = J_z(0)e^{i(\mathbf{k}\cdot{r}-\omega t)}$. We shall write $\bar{\omega} = \omega_s + i\gamma_0$, so that 
\begin{equation}
\gamma = 1/\tau - \gamma_0 \label{gam}
\end{equation}
by definition. We shall see that $\gamma_0$ is negative in all models except for the SCM, in which it is positive-definite thanks to a fundamental physical reason. 

For linear responses, $\mathbf{J}(z)$ can be related to $\mathbf{E}(z)$ as follows 
\begin{equation}
J_\mu(z) = \sum_{\mu,\nu = x,y,z} \int dz' \sigma_{\mu\nu}(z,z',\omega)E_\nu(z'),
\end{equation}
where $\sigma_{\mu\nu}(z,z',\omega)$ is the conductivity tensor. Considering that $\mathbf{E}(z)$ linearly depends on $\rho(z)$, we can define a linear operator $\hat{\mathcal{H}}$ with kernel $\mathcal{H}(z,z')$ so that
\begin{equation}
\hat{\mathcal{H}}\rho(z) = \int dz' \mathcal{H}(z,z') \rho(z') = -i\bar{\omega}\nabla\cdot\mathbf{J}(z). \label{2.4}
\end{equation}
It is easy to show that
\begin{equation}
\mathcal{H}(z,z') = i\bar{\omega}\sum_{\mu\nu}\int^\infty_0 dz'' \nabla_\mu\sigma_{\mu\nu}(z,z'',\omega)\nabla''_\nu V_\mathbf{k}(z''-z'), \nonumber
\end{equation}
where $\nabla'' = (i\mathbf{k},\partial_{z''})$ and $V_\mathbf{k}(z)$ is the $\mathbf{k}$-th Fourier component of the Coulomb interaction $V(\mathbf{x}) = 1/\abs{\mathbf{x}}$. Now Eq.~(\ref{2.2}) can be transformed into 
\begin{equation}
\left(\hat{\mathcal{H}}-\bar{\omega}^2\right)\rho(z) = S\Theta'(z), \quad S = i\bar{\omega}J_z(0).  \label{2.5}
\end{equation}
Note that $S$ does not depend on $z$. Since $\rho(z)$ is defined only for $z\geq0$, we can introduce a cosine Fourier transform $$\rho(z) = \frac{2}{\pi}\int^\infty_0 dq ~ \rho_q\cos(qz).$$ In terms of $\rho_q$, Eq.~(\ref{2.5}) can be rewritten as
\begin{equation}
\int^\infty_0 dq' \left(\mathcal{H}(q,q') - \bar{\omega}^2\delta(q-q')\right)\rho_{q'} = S. \label{2.6}
\end{equation}
Here $$\mathcal{H}(q,q') = \frac{2}{\pi}\int^\infty_0dz\int^\infty_0dz'\cos(qz)\mathcal{H}(z,z')\cos(q'z')$$ is the matrix element between the cosine waves. Finally, we close Eq.~(\ref{2.6}) by the fact that $J_z(0)$ and hence $S$ are also linear functionals of $\rho_q$. We can thus write
\begin{equation}
S = \int^\infty_0 dq \frac{G(\mathbf{K},\omega)}{K^2} \rho_q, \label{2.7}
\end{equation}
where $\mathbf{K} = (k,0,q)$, $K^2 = k^2+q^2$ and $G$ is a kernel given by $$G(\mathbf{K},\omega) = \frac{2\bar{\omega}}{i\pi}\int^\infty_0dz\int^\infty_0dz' \cos(qz)\sigma_{\mu\nu}(0,z',\omega)\nabla'_\nu V_\mathbf{k}(z'-z).$$ As a key result of this paper, Eqs.~(\ref{2.6}) and (\ref{2.7}) constitute a complete description of self-sustained charge density waves in SIMs. Their basic structures are valid regardless of the underlying electron dynamics encoded in $\mathcal{H}$ and $G$.  

\subsection{Universal secular equation for SPWs}
\label{sec:2.3}
Equation~(\ref{2.6}) is an inhomogeneous linear equation with a source term. Equations of this type generally admit of two classes of solutions, depending on whether $S$ vanishes or not. Solutions with $S = 0$ represent nothing but the bulk plasma waves, while those with $S\neq 0$ represent localized waves, i.e. SPWs in our system. The spectra of these two classes generally do not overlap. Had we imposed auxiliary boundary conditions as is usually done in the literature, SPWs would not exist at all.  

To find the 'secular' equation for SPWs, we wish to make a simplification, which is not necessary but will make the resulting expressions more transparent. To this end, let us look at some general properties of $\sigma_{\mu\nu}(z,z',\omega)$. For an infinite system without boundaries, the translational symmetry is preserved along $z$-axis as well as along the surface plane, in which case $\sigma_{\mu\nu}$ only depends on $z-z'$. For SIMs, however, that symmetry is broken and $\sigma_{\mu\nu}$ must in general depend on $z$ and $z'$ separately. It shall prove useful to decompose $\sigma_{\mu\nu}$ into two parts, $\sigma_{b,\mu\nu}$ and $\sigma_{s,\mu\nu}$, where $\sigma_{b,\mu\nu}$ is defined to be that of the infinite system and depends only on $z-z'$ while $\sigma_{s,\mu\nu}$ signifies pure surface effects. Namely, $$\sigma_{\mu\nu}(z,z',\omega) = \sigma_{b,\mu\nu}(z-z',\omega) + \sigma_{s,\mu\nu}(z,z',\omega).$$ Accordingly, $\mathcal{H}$ and $G$ also each contain two parts arising from $\sigma_{b,\mu\nu}$ and $\sigma_{s,\mu\nu}$, respectively. Let us then write $\mathcal{H} = \mathcal{H}_b + \mathcal{H}_s$. Now that it reflects on the properties of plasma waves of an infinite system, $\mathcal{H}_b$ must be diagonal in the $q$-space, i.e. $\mathcal{H}_b(q,q') = \Omega^2(K,\omega)\delta(q-q')$, where $\Omega(K,\omega)$ is a frequency that only depends on $K$. By definition, one can easily show that the dielectric function of an infinite system is given by
\begin{equation}
    \epsilon(K,\omega) = 1 - \frac{\Omega^2(K,\omega)}{\bar{\omega}^2}.
\end{equation}
As for $\mathcal{H}_s$, it gives rise to scattering of plasma waves. Nevertheless, for bulk waves the scattering due to a surface should be insignificant and may be treated perturbatively, as we did in Refs.~\cite{deng2017a,deng2017b}. To the lowest order in this perturbation, we may simply put
\begin{equation}
\mathcal{H}(q,q') = \Omega^2(K,\omega) \delta(q-q'). \label{2.8}
\end{equation}
This expression contains complete information of bulk plasma waves. As expected, the zeros of $\epsilon(K,\omega)$ give the dispersion $\epsilon_b(K)$ of these waves. 

We analogously write $G = G_b + G_s$. Note that via Eq.~(\ref{2.7}) $G_b$ determines $J_z$ at the point $z=0$ in a infinite system, for which, however this point is none more special than any other points. Thus, $G_b$ can only be a constant that does not depend on $q$ and it is therefore purely due to the local part of $\sigma_{b,\mu\nu}(z-z')$. This part must be isotropic for the jellium model and it can be written as $\delta_{\mu\nu}\delta(z-z')\sigma(\omega)$, where $\delta_{\mu\nu}$ is the Kroneckle symbol and $\sigma(\omega)$ is to be discussed further. It follows that $G_b = -4i\bar{\omega}\sigma(\omega)k\beta$, where we have taken into account the effect of the dielectric via the factor $\beta = 2\epsilon_d/(1+\epsilon_d)$; see details below. We now obtain
\begin{equation}
G(\mathbf{K},\omega) = -4i\bar{\omega}\sigma(\omega)k\beta + G_s(\mathbf{K},\omega). \label{2.9}
\end{equation}
We see that $G_s$ describes translation symmetry breaking effects. As to be seen later, it is missing from the DM, the HDM and the SRM. The positive-definiteness of $\gamma_0$ arises solely through this term and therefore cannot be captured in these models. 

For SPWs, $S\neq0$. Inserting Eq.~(\ref{2.8}) in (\ref{2.6}) and using (\ref{2.7}), we find
\begin{equation}
\epsilon_s(k,\omega) = 1 - \int^\infty_0 \frac{dq}{K^2}\frac{G(\mathbf{K},\omega)}{\Omega^2(K,\omega)-\bar{\omega}^2} = 0,\label{2.11}
\end{equation}
which determines the SPW frequency $\omega_s$ and damping rate $\gamma$ as a function of $k$. This is another key result of this paper. The SPW charge density is obtained as
\begin{equation}
    \rho_q = \frac{S}{\Omega^2(K,\omega) - \bar{\omega}^2} = \frac{S}{\bar{\omega}^2}\left(-\frac{1}{\epsilon(K,\omega)}\right). \label{2.12}
\end{equation}
The resulting $\rho(z)$ is peaked about the surface. 

Equations (\ref{2.11}) and (\ref{2.12}) constitute a universal description of SPWs. They are valid irrespective of the underlying electron dynamics, which enter only through $\epsilon(K,\omega)$ and $G(\mathbf{K},\omega)$. 

\subsection{Dielectric effects}
\label{sec:2.4}
Let us place a dielectric with dielectric constant $\epsilon_d$ -- which can be complex -- on the side $z<0$. The electrostatic fields are affected by this dielectric, which can be calculated with the method of mirror charges. Let the mirror charge density in the dielectric be $\rho_d(\mathbf{x},t) = \rho_d(z)e^{i(\mathbf{k}\cdot\mathbf{r}-\omega t)}$. It is easy to show that~\cite{pendry1975}
\begin{equation}
    \rho_d(z<0) = - \frac{\epsilon_d-1}{\epsilon_d+1} \rho(-z). 
\end{equation}
The electrostatic potential $\phi(\mathbf{x},t) = \phi(z)e^{i(\mathbf{k}\cdot\mathbf{r}-\omega t)}$, obeying Poisson's equation $\partial^2_\mathbf{x}\phi(\mathbf{x},t) + 4\pi(\rho(\mathbf{x},t) + \rho_d(\mathbf{x},t)) = 0$, is then obtained as
\begin{equation}
    \phi(z) = \frac{2\pi}{k}\int^\infty_{-\infty} dz' ~ e^{-k\abs{z-z'}}\left(\rho(z')+\rho_d(z')\right).
\end{equation}
In the metal, where $z\geq 0$, the electric field $\mathbf{E}(z) = - \nabla \phi(z) = (E_x(z),0,E_z(z))$ then follows as
\begin{equation}
    E_x(z) = -i\int^\infty_0dq ~ \frac{4k \rho_q}{K^2}\left(2\cos(qz) - \beta e^{-kz}\right) \label{2.15}
\end{equation}
and 
\begin{equation}
    E_z(z) = \int^\infty_0dq ~ \frac{4k \rho_q}{K^2}\left(2\frac{q}{k}\sin(qz) - \beta e^{-kz}\right), \label{2.16}
\end{equation}
where $\beta$ has been given in the above. See that $E_z(0)$ and hence $G_b$ are enhanced by the factor $\beta$, which is why this factor appears in Eq.~(\ref{2.9}). Physically this is because the surface sits between the charges in the metal and those in the dielectric and spatially separates them. 

\section{Inter-band transition effects}
\label{sec:2.5}
The dynamics models to be considered in the next two sections only describe the currents from electrons in the conduction band. Realistically, valence electrons can also contribute by virtual inter-band transitions. In this paper, for the sake of simplicity, we account for these inter-band transition effects by a phenomenological approach~\cite{liebsch1993,rakic1998,rakic1995}. The observation is that, valence electrons are usually tightly held to their host atoms and the energy bands are largely non-dispersion, and thus their electrical responses are mostly local and not susceptible to the presence of boundaries. We may then describe this response by a local conductivity function $\sigma_p(\omega)$, so that the electrical current density due to the valence electrons is given by 
\begin{equation}
\mathbf{J}_p(z) = \sigma_p(\omega) \mathbf{E}(z).    
\end{equation}
Usually $\sigma_p(\omega)$ may be modeled in the Lorentz form. It is related to the inter-band dielectric function by $\epsilon_p(\omega) = 4\pi i\sigma_p(\omega)/\bar{\omega}$, which can be measured for example by means of ellipsometry or computed by density functional theory. $\epsilon_p(\omega)$ contains a real part $\epsilon_{pr}(\omega)$ and an imaginary part $\epsilon_{pi}(\omega)$. While $\epsilon_{pr}(\omega)$ acts to shield the conduction electrons, $\epsilon_{pi}(\omega)$ -- which is always positive -- leads to inter-band absorption. Being basically an atomic property, $\epsilon_p$ is not sensitive to temperature.

\section{SPWs in DM, HDM and SRM}
\label{sec:3}
While the theory established in Sec.~\ref{sec:2} is universal, the behaviors of SPWs do depend on electron dynamics through $\epsilon(K,\omega)$ and $G(\mathbf{K},\omega)$. In this section, we apply the theory to examine SPWs within several common electron dynamics models: the DM, the HDM and the SRM. SPWs in the SCM will be thoroughly treated in the next section. We show that the usually quoted SPW solution in the HDM is false and clarify the origin of SPWs in the SRM. A summary of this analysis is tabulated in Table~\ref{t:1}

\begin{table*}
\caption{\label{t:1} A comparison of various models in light of the present universal theory of SPWs. DM: classical dielectric model. HDM: hydrodynamic model. SRM: specular reflection model. SCM: semi-classical model. For the sake of simplicity, $\epsilon_p = 0$ and $\epsilon_d = 1$ have been used in the expressions given in this table. $\mathbf{K} = (k,0,q)$. $G_0 = (k/\pi)\omega^2_p$, where $\omega_p$ denotes the characteristic plasma frequency of the metal. $\mathbf{F} = (m/2\pi\hbar)^3\int d^3\mathbf{v}(-e^2f'_0)\mathbf{v}(\mathbf{K}\cdot\mathbf{v}/\bar{\omega})^2(1-\mathbf{K}\cdot\mathbf{v}/\bar{\omega})^{-1}$, where $\mathbf{v}$ is the velocity of electrons of mass $m$. $\epsilon(\mathbf{K},\omega)$ is the dielectric function of the infinite system, whose zeros give the dispersion of bulk plasma waves. SPWs are determined by this secular equation: $\epsilon_s = 1 + \int^\infty_0 \frac{dq}{K^2}\frac{G(\mathbf{K},\omega)}{\bar{\omega}^2}\frac{1}{\epsilon(\mathbf{K},\omega)} = 0$. The solution to this equation is written as $\omega = \omega_s - i\gamma$. In the SCM, an extra contribution $\gamma_s$ arises due to symmetry breaking effects contained in $G_s$. Very interestingly, $\gamma_0 = \gamma_s - \gamma_\text{Landau} - \gamma_\text{interband}$ cannot be negative.}
\begin{ruledtabular}
\begin{tabular}{c c c c c}
Quantity & DM & HDM & SRM & SCM \\ \hline
Symmetry breaking effects & No & No & No & Yes \\
$\Omega^2(K,\omega)$ & $\omega^2_p$ & $\Omega^2_\text{HDM} = \omega^2_p + Bv^2_FK^2$ & $\omega^2_p + \frac{4\pi\bar{\omega}\mathbf{K}\cdot\mathbf{F}}{K^2}$ & $\omega^2_p + \frac{4\pi\bar{\omega}\mathbf{K}\cdot\mathbf{F}}{K^2}$\\
$G(\mathbf{K},\omega)$ & $G_0$ & $G_0$ & $G_0$ & $G_0 + G_s$ \\
$\epsilon(\mathbf{K},\omega)$ & $1-\frac{\omega^2_p}{\bar{\omega}^2}$ & $1-\frac{\Omega^2_\text{HDM}}{\bar{\omega}^2}$ & $1-\frac{\Omega^2}{\bar{\omega}^2}$ & $1-\frac{\Omega^2}{\bar{\omega}^2}$ \\
Damping rate $\gamma$ & $\tau^{-1}+\gamma_\text{interband}$ & $\tau^{-1}+\gamma_\text{interband}$ & $\tau^{-1}+\gamma_\text{interband} + \gamma_\text{Landau}$ & $\tau^{-1}+\gamma_\text{interband} + \gamma_\text{Landau} - \gamma_s$
\end{tabular}
\end{ruledtabular}
\end{table*}

\subsection{Local dielectric model (DM)}
\label{sec:3.1}
This is the standard model for SPWs. Unlike other models, it does not require and is incompatible with any auxiliary conditions. Here we reproduce by our theory the well-known properties of SPWs in this model. 

According to the DM, the current density due to conduction electrons is given by
\begin{equation}
    \mathbf{J}_\text{DM}(z) = \sigma_\text{DM}(\omega)\mathbf{E}(z), \quad \sigma_\text{DM}(\omega) = \frac{i}{\bar{\omega}}\frac{\omega^2_p}{4\pi}, 
\end{equation}
where $\omega_p = \sqrt{4\pi n_0e^2/m}$ is the characteristic plasma frequency of the metal with $m$ and $e$ being the effective mass and charge of an electron, respectively. The total current density $\mathbf{J}(z)$ then is $$\mathbf{J}_p(z) + \mathbf{J}_\text{DM}(z) = (\sigma_\text{DM}(\omega)+\sigma_p(\omega))\mathbf{E}(z),$$ yielding $\sigma(\omega) = \sigma_\text{DM}(\omega)+\sigma_p(\omega)$. From Eq.~(\ref{2.4}) one finds $\Omega^2$ dispersionless, given as 
$$\Omega^2_0(\omega) = \omega^2_p - \bar{\omega}^2\epsilon_p(\omega)$$ 

\textit{Bulk plasma waves}. Equating $\Omega^2_0$ with $\bar{\omega}^2$ yields the frequency $\omega_\text{b,DM}$ and damping rate $\gamma_\text{b,DM}$ for bulk plasma waves. For $\omega_p\tau\rightarrow \infty$ and assuming $\epsilon_p(\omega)$ independent of $\omega$, they are given by $$\omega_\text{b,DM} = \frac{\omega_p}{\sqrt{1+\epsilon_{pr}}}, \quad \gamma_\text{b,DM}/\omega_\text{b,DM} = \frac{1}{\omega_\text{b,DM}\tau} + \frac{1}{2}\frac{\epsilon_{pi}}{1+\epsilon_{pr}}.$$ Note that bulk waves bear no dielectric effects, i.e. no dependence on $\epsilon_d$. The damping rate $\gamma_\text{b,DM}$ arises due to thermal collisions and inter-band absorption.  

\textit{SPWs}. The electrical conductivity is purely local and thus $G_s = 0$. As such, $G(\mathbf{K},\omega)$ becomes $$G_0 = (k\beta/\pi)\Omega^2_0(\omega).$$ Substituting this into Eq.~(\ref{2.11}), we immediately arrive at the often quoted frequency $\omega_\text{s,DM}$ and damping rate $\gamma_\text{s,DM}$ for SPWs. Neglecting absorption in the dielectric, i.e. assuming real $\epsilon_d$, we find
$$\omega_\text{s,DM} = \frac{\omega_p}{\sqrt{1+\epsilon_d+\epsilon_{pr}}}, \quad \frac{\gamma_\text{s,DM}}{\omega_\text{s,DM}} = \frac{1}{\omega_\text{s,DM}\tau} + \frac{1}{2}\frac{\epsilon_{pi}}{1+\epsilon_d+\epsilon_{pr}}.$$ In this model, the SPW charge density is completely localized on the surface, $\rho(z) = \rho_s\Theta'(z)$, where $\rho_s = S/(\Omega^2_0 - \bar{\omega}^2)$ gives the areal charge density. See that the dielectric tends to reduce the SPW damping rate. 

Traditionally~\cite{raether1988}, the above results have been obtained by treating the metal as a simple dielectric with dielectric constant $\epsilon(\omega) = 1 - \Omega^2_0/\bar{\omega}^2$. Then exponentially decaying electromagnetic (EM) waves (or electrostatic potentials in the quasi-static limit) are written down on the metal and the dielectric sides, and Maxwell's boundary conditions are used to match the waves to obtain the above frequency and damping rate of SPWs. Our theory works directly with charge density rather than EM waves. The two approaches are equivalent. 

\subsection{Hydrodynamic model (HDM)}
\label{sec:3.2}
The DM assumes a purely local relation between the current density and the electric field. The HDM extends the DM by inclusion of leading-order non-local corrections. Recently, this model has attracted lots of attention in plasmonics and quantum forces~\cite{pendry2013,luo2014}. It has also been synergized with density functional theory in the quantum hydrodynamic model~\cite{toscano2015,yan2015,ciraci2016,christensen2017} to study local plasmon resonances on metal particles. 

In the HDM, $\Omega^2$ includes leading-order dependence on $K$ and is given by $$\Omega^2_\text{HDM} = \Omega^2_0 + K^2v^2_0,$$ where $v_0$ is a parameter. The dielectric function~\cite{notehdm} then reads $\epsilon_\text{HDM} = 1 - \Omega^2_\text{HDM}/\bar{\omega}^2$. The bulk waves are similar to those in the DM except for some dispersion with $K$. 

As in the DM, no symmetry breaking effects are included in the HDM. $G(\mathbf{K},\omega)$ is then the same as that for the DM. Now Eq.~(\ref{2.11}) transforms into the following
\begin{equation}
1+\frac{k\beta}{\pi} \frac{\Omega^2_0}{\bar{\omega}^2}\int^{\infty}_0 \frac{dq}{K^2} \frac{1}{\epsilon_\text{HDM}} = 0, \label{2.181} 
\end{equation}
which determines the SPWs in the HDM by our theory. Note that the integrand contains no poles or resonances near the solutions, as the SPW spectra are always gapped from the bulk wave spectra gratifying $\epsilon_\text{HDM} = 0$. For $v_0=0$, the HDM reduces to the DM and so do the SPWs, as expected. With $\epsilon_d=0$ and $\epsilon_p=0$, the equation simplifies to
\begin{equation}
    1+\frac{k}{\pi} \int^{\infty}_{-\infty} \frac{dq}{K^2} \frac{1}{\epsilon_\text{HDM}} = 0, \label{2.18}
\end{equation}
where we have used the fact that $\epsilon_\text{HDM}$ is even in $q$ and that $\omega^2_p/\bar{\omega}^2\approx 2$ for the solutions to this equation. As shown in Appendix~\ref{sec:a1}, it leads to a linear dispersion of $\omega_s$ versus $k$. As in the DM, the SPW damps due to thermal collisions and inter-band absorption, at rate $\gamma_\text{s,HDM} \approx \gamma_\text{s,DM}$ apart from some dispersion effects. 

In the literature, the condition that $J_z(0) = 0$ is usually imposed in the HDM~\cite{harris1971,garcia1977b}. This would mean $S=0$ and therefore would exclude any SPWs according to our theory. Nevertheless, SPWs have been claimed to exist under this condition. In what follows we briefly show that this claim is false, more details to be found in Appendix~\ref{sec:a1}. 

For illustration, we take $\epsilon_p=0$ and $\epsilon_d=0$. Impose $S=0$ and the wave equation becomes $\left(\Omega^2_\text{HDM}-\bar{\omega}^2\right)\rho_q = 0$, or equivalently in the real space
\begin{equation}
    \left(\bar{\omega}^2-(\omega^2_p + v^2_0k^2) + v^2_0\partial^2_z\right)\rho(z) = 0. \label{2.20}
\end{equation}
The claimed SPW solution is then sought of the form $\rho(z) = \rho_0e^{-\kappa z}$. Substituting this in the equation leads to $\bar{\omega}^2 = \omega^2_p + v^2_0(k^2-\kappa^2)$. Imposing $J_z(0) = 0$ gives another relation, $\omega^2_{0} = v^2_0\kappa(k+\kappa)$, which expresses the balance between the electronic pressure and the electric force at the surface. Here $\omega_{0} = \omega_p/\sqrt{2}$. Those two relations specify the solution. Combined, they lead to $\bar{\omega}^2 \approx \omega^2_{0} + \omega_{0}\beta k$. Nevertheless, this solution does not reduce in the limit $v_0=0$ to the SPWs found by Ritchie with the DM. Actually, $\kappa$ diverges in this limit, yielding $\int \rho(z) dz \sim \kappa^{-1} = 0$, i.e. the solution is empty of charges. This false solution is also what was observed in Refs.~\cite{harris1972,garcia1977b,feibelman1971}. It is plausible that existing \textit{ab initio} calculations have only observed this false solution as well~\cite{tsuei1991}. A comprehensive account may merit a future study. 

Although the false solution and the correct solution are conceptually disparate, their dispersion relations are quite similar, as shown in Appendix~\ref{sec:a1}. 

\subsection{Specular reflection model (SRM)}
\label{sec:3.3}
A natural step to go beyond the HDM is to use the full form of $\Omega$ rather than the approximation $\Omega_\text{HDM}$.  Equation (\ref{2.18}) then becomes
\begin{equation}
    1+\frac{k}{\pi} \int^{\infty}_{-\infty} \frac{dq}{K^2} \frac{1}{\epsilon(\mathbf{K},\omega)} = 0. \label{2.19}
\end{equation}
This is exactly the equation established by Marusak and Ritchie in 1966 for the SRM~\cite{ritchie1966}. Our derivation makes it clear that the SRM is just an extension of the HDM. From this point of view, one may also conclude that the usually claimed SPWs in the HDM are false, because they are not solutions of Eq.~\ref{2.19} in the HDM limit. 

In contrast with the DM and the HDM, SPWs in the SRM can also decay via Landau damping, because of an imaginary part in $\Omega$ associated with electron-hole excitations. Thus, the SPW damping rate is $\gamma_\text{s,SRM} \approx \gamma_\text{s,DM} + \gamma_\text{Landau}$, see the next section for further discussion on this. 

We wish to point out a logical inconsistency in the original contrivance of the SRM. There are two elements in this contrivance: (i) as nominally expected, electron waves impinging on the surface are assumed to be specularly reflected back, and (ii) a sheet of 'fictitious' charges exactly localized on the surface. Element (i) would mean $J_z(0)=0$ and hence, by our theory, no SPWs would materialize. Then how do those waves come about? The answer rests with element (ii). In Appendix~\ref{sec:a2}, we show that the fictitious charge sheet reinstates the capacitive effects lost under element (i). Actually, SPWs appear as a pole of this fictitious charge density. 

As with the HDM and the DM, the SRM contains no symmetry breaking effects, i.e. $G_s$ is absent from these models. To account for these effects, further improvement is required, leading to the SCM. 
 
\section{SPWs in the SCM}
\label{sec:4}
In the SCM one calculates the electric currents in terms of a distribution function $f(\mathbf{x},\mathbf{v},t)$ defined in the single-particle phase space. As usual, we write it as a sum of the equilibrium part $f_0(\varepsilon(\mathbf{v}))$ and the non-equilibrium part $g(\mathbf{x},\mathbf{v},t)$. $f_0(\varepsilon)$ is taken to be the Fermi-Dirac function at zero temperature. $\varepsilon(\mathbf{v}) = mv^2/2$ is the dispersion of the conduction band. Within the relaxation time approximation and the regime of linear responses, $g(\mathbf{x},\mathbf{v},t) = g(\mathbf{v},z) e^{i(\mathbf{k}\cdot\mathbf{r}-\omega t)}$ satisfies the following Boltzmann's equation
\begin{equation}
    \left(\lambda^{-1} + \partial_z\right)g(\mathbf{v},z) + ef'_0(\varepsilon)\mathbf{v}\cdot\mathbf{E}(z)/v_z = 0. \label{3.1}
\end{equation}
Here $\lambda = iv_z/\tilde{\omega}$ with $\tilde{\omega} = \bar{\omega} - kv_x$ and $f'_0 = \partial_{\varepsilon}f_0(\varepsilon) = (2/m)\delta(v^2-v^2_F)$, where $m$ is the electron effective mass and $v_F$ the Fermi velocity. 

Physical causality~\cite{deng2017a} requires that $\gamma_0 = \text{Im}(\bar{\omega})\geq0$; otherwise, reflected electron waves would come before incident waves. Together with Eq.~(\ref{gam}), we may conclude that the SPW damping rate is always in short of $\tau^{-1}$, i.e. $\gamma \tau <1$, in non-reconcilable contrast with other models. 

With $\gamma_0\geq 0$, the general solution to Eq.~(\ref{3.1}) can be written as 
\begin{equation}
g(\mathbf{v},z) = e^{-\frac{z}{\lambda}}\left(C(\mathbf{v})-\frac{ef'_0\mathbf{v}}{v_z}\cdot \int^z_0~dz'~e^{\frac{z'}{\lambda}}~\mathbf{E}(z')\right), \label{3.2}
\end{equation}
where $C(\mathbf{v}) = g(\mathbf{v},0)$ is the non-equilibrium deviation on the surface to be determined by boundary conditions. We require $g(\mathbf{v},z)=0$ distant from the surface, i.e. $z\rightarrow\infty$. For electrons moving away from the surface, $v_z>0$, this condition is automatically fulfilled. For electrons moving toward the surface, $v_z<0$, it leads to
\begin{equation}
C(\mathbf{v}) =\frac{ef'_0\mathbf{v}}{v_z}\cdot \int^{\infty}_0~dz'~e^{z'/\lambda}\mathbf{E}(z'), \quad v_z<0.
\end{equation}
It follows that
\begin{equation}
g(\mathbf{v},z) = \frac{ef'_0\mathbf{v}}{v_z}\cdot \int^{\infty}_zdz' ~e^{\frac{z'-z}{\lambda}}~\mathbf{E}(z'), \quad v_z<0. 
\end{equation}
To determine $C(\mathbf{v})$ for $v_z>0$, the boundary condition at $z=0$ has to be used, which, whoever, depends on surface properties. We adopt a simple picture first conceived by Fuchs~\cite{fuchs1938} and then widely used in the study of for instance anomalous skin effect~\cite{ziman,abrikosov,reuter1948,kaganov1997}. According to this picture a fraction $p$ i.e. the \textit{Fuchs} parameter varying between zero and unity, of the electrons impinging on the surface are specularly reflected back, i.e. 
\begin{equation}
g(\mathbf{v},z=0)=p~g(\mathbf{v}_-,z=0), \quad \mathbf{v}_- = (v_x,v_y,-v_z), \quad v_z\geq 0. \label{3.5} 
\end{equation}
This condition is identical with the condition used in Ref.~\cite{deng2017c} but differs from that in Refs.~\cite{deng2017a,deng2017b} except for $p=0$. It follows that 
\begin{equation}
C(\mathbf{v}) = - p~\frac{ef'_0\mathbf{v}_-}{v_z}\cdot \int^{\infty}_0dz' ~e^{-\frac{z'}{\lambda}}~\mathbf{E}(z'), \quad v_z\geq0. \label{3.6}
\end{equation} 
Equations (\ref{3.2}) - (\ref{3.6}) fully specify the distribution function for the electrons. 

The electrical current density due to the conduction electrons is then calculated as
\begin{equation}
\mathbf{J}_c(z) = (m/2\pi\hbar)^3\int d^3\mathbf{v} ~e\mathbf{v}g(\mathbf{v},z). \label{3.7} 
\end{equation}
Note that the charge density is not given by $$\tilde{\rho}(\mathbf{x},t)=(m/2\pi\hbar)^3~e^{i(kx-\omega t)}\int d^3\mathbf{v}~eg(\mathbf{v},z).$$ The reason is simple: the as-obtained $g(\mathbf{v},z)$ is for the bulk region and not valid on the surface, because Eq.~(3.1) involves no surface potentials, see Sec.~\ref{sec:2.1}. Actually, $\mathbf{J}_c(\mathbf{x},t)$ and $\tilde{\rho}(\mathbf{x},t)$ obey the equation $$(\partial_t+1/\tau)\tilde{\rho}(\mathbf{x},t)+\partial_{\mathbf{x}}\cdot\mathbf{J}_c(\mathbf{x},t)=0$$ rather than the equation of continuity [c.f. Eq.~(\ref{2.1})], automatically embodying the condition that $J_z(0)=0$. This underlies the incorrect conclusion drawn by Harris and others~\cite{harris1972}. 

\subsection{The positive-definiteness of $\gamma_0$}
\label{sec:4.1}
We substitute the expression of $\mathbf{E}(z)$ given by Eqs.~(\ref{2.15}) and (\ref{2.16}) into (\ref{3.2}) - (\ref{3.6}) and perform the integration over $z'$. We find it instructive to split $g(\mathbf{v},z)$ into two parts, one denoted by $g_b(\mathbf{v},z)$ and the other by $g_s(\mathbf{v},z)$. They are given by
\begin{eqnarray}
g_b(\mathbf{v},z) &=& -ef'_0\int^\infty_0 dq \frac{4\rho_q}{K^2}\times \nonumber \\&~& \quad \left(2F_+\cos(qz)+2iF_-\sin(qz)-\beta F_0e^{-kz}\right), \label{gb}
\end{eqnarray}
where we have introduced the following functions,
\begin{equation}
F_\pm(\mathbf{K},\bar{\omega},\mathbf{v}) = \frac{1}{2}\left[\frac{\mathbf{K}\cdot\mathbf{v}}{\bar{\omega} - \mathbf{K}\cdot\mathbf{v}}\pm \frac{\mathbf{K}\cdot\mathbf{v}_-}{\bar{\omega} - \mathbf{K}\cdot\mathbf{v}_-}\right].
\end{equation}
Note that $F_{+/-}$ is an even/odd function of $v_z$. In addition, 
\begin{equation}
F_0(\mathbf{k},\bar{\omega},\mathbf{v}) = \frac{\mathbf{k}^*\cdot\mathbf{v}}{\bar{\omega} - \mathbf{k}^*\cdot\mathbf{v}} = \sum^\infty_{l=1}\left(\frac{\mathbf{k}^*\cdot\mathbf{v}}{\bar{\omega}}\right)^l, \quad \mathbf{k}^* = (k,0,ik). \end{equation}
Moreover, we have
\begin{eqnarray}
&~&g_s(\mathbf{v},z) = \Theta(v_z)(-ef'_0)e^{i\frac{\bar{\omega}z}{v_z}}\int^\infty_0 dq\frac{4\rho_q}{K^2}\times \label{gs} \\ &~& \quad \left[\beta F_0(\mathbf{k},\bar{\omega},\mathbf{v})-p\beta F_0(\mathbf{k},\bar{\omega},\mathbf{v}_-) + 2(p-1)F_+(\mathbf{K},\bar{\omega},\mathbf{v})\right]. \nonumber
\end{eqnarray}
One may show that $g_b(\mathbf{v},z)$ can also be obtained directly by Boltzmann's equation for an infinite system. Thus, this part contains exactly the responses of an infinite system. It is independent of surface properties, i.e. showing no dependence on the \textit{Fuchs} parameter $p$, and the electrons incident on the surface (i.e. with $v_z<0$) and those departing it (i.e. with $v_z>0$) appear on equal footing in its expression. 

On the other hand, $g_s(\mathbf{v},z)$ signifies pure surface effects: it exists only for departing electrons, as indicated by the Heaviside function $\Theta(v_z)$ in its expression, and it depends on $p$ thus reflecting on surface roughness. If we keep only $g_b(\mathbf{v},z)$, the SRM will be revisited, making it evident that the SRM does not correspond to the limit of $p=1$ (specularly reflecting surface), in contrast with its nominal meaning.  

Another important feature of $g_s(\mathbf{v},z)$ lies in its dependence on $z$, i.e. $g_s(\mathbf{v},z) \propto e^{i\tilde{\omega}z/v_z} \propto e^{-\gamma_0z/v_z}$, which implies that $\gamma_0\geq0$ in accord with causality [see also preceding remarks above Eq. (\ref{3.2})]. Otherwise, it would diverge far away from the surface. 

\subsection{$\Omega$ and $G$ in the SCM}
\label{sec:4.2}
The conduction current density is also written in two parts, $\mathbf{J}_c(z) = \mathbf{J}_b(z) + \mathbf{J}_s(z)$, where $\mathbf{J}_{b/s}(z)$ are defined via Eq.~(\ref{3.7}) with $g(\mathbf{v},z)$ replaced by $g_{b/s}(\mathbf{v},z)$. For small $kv_F/\bar{\omega}$, we may keep only the first term in the series of $F_0(\mathbf{k},\bar{\omega},\mathbf{v})$. It is then straightforward to show that 
\begin{equation}
    \mathbf{J}_b(z) = \sigma_\text{DM}(\omega) \mathbf{E}(z) + \mathbf{J}_\text{SRM}(z),
\end{equation}
where $\mathbf{J}_\text{SRM}(z)$ is responsible for the extension made through the SRM beyond the DM. It is given by~\cite{note3.1}
\begin{eqnarray}
J_{\text{SRM},x}(z) &=& \int \mathcal{D}q\mathcal{D}^3\mathbf{v}~v_x F'_+(\mathbf{K},\bar{\omega},\mathbf{v})\cos(qz), \\
J_{\text{SRM},z}(z) &=& i ~ \int \mathcal{D}q\mathcal{D}^3\mathbf{v}~v_z F'_-(\mathbf{K},\bar{\omega},\mathbf{v})\sin(qz),
\end{eqnarray}
where we have defined a short-hand
$$\int \mathcal{D}q\mathcal{D}^3\mathbf{v} ... = \left(\frac{m}{2\pi\hbar}\right)^3 \int^\infty_0 dq~\frac{4\rho_q}{K^2} \int d^3\mathbf{v}\left(-e^2f'_0\right) ...$$
together with these functions $$F'_\pm(\mathbf{K},\bar{\omega},\mathbf{v}) = \frac{1}{2}\left[\frac{(\mathbf{K}\cdot\mathbf{v})^2}{1 - \mathbf{K}\cdot\mathbf{v}/\bar{\omega}}\pm \frac{(\mathbf{K}\cdot\mathbf{v}_-)^2}{1 - \mathbf{K}\cdot\mathbf{v}_-/\bar{\omega}}\right].$$ See that $J_{\text{SRM},z}(0)\equiv 0$, which makes no contribution to $G$. Now the total current density becomes $$\mathbf{J}(z) = \mathbf{J}_b(z) + \mathbf{J}_s(z) + \mathbf{J}_p(z).$$
By definitions (\ref{2.4}) and (\ref{2.8}), we find
\begin{equation}
    \Omega^2(K,\omega) = \Omega^2_0(\omega) + \frac{4\pi\bar{\omega}\mathbf{K}\cdot\mathbf{F}(\mathbf{K},\bar{\omega})}{K^2}, \label{3.15}
\end{equation}
where $\mathbf{F}$ is an odd function of $\bar{\omega}$ and given by
\begin{equation}
    \mathbf{F}(\mathbf{K},\bar{\omega}) = \left(\frac{m}{2\pi\hbar}\right)^3 \int d^3\mathbf{v}\left(-e^2f'_0\right)\mathbf{v}\frac{\left(\mathbf{K}\cdot\mathbf{v}/\bar{\omega}\right)^2}{1-\mathbf{K}\cdot\mathbf{v}/\bar{\omega}}. 
\end{equation}
Partially performing the integral, we obtain
\begin{equation}
\Omega^2(K,\omega) = \omega^2_p\left(1 - \frac{\bar{\omega}^2\epsilon_p}{\omega^2_p} + \frac{3}{2}\frac{Kv_F}{\bar{\omega}}\int^1_{-1} dr\frac{r^3}{1-rKv_F/\bar{\omega}}\right).
\end{equation}
The real part of this expression approximates $\omega^2_p\left(1+\frac{3}{5}K^2/k^2_p\right)$ in the long wavelength limit, corresponding to the HDM limit with $v_0 = \sqrt{3/5}v_F$. Here $k_p = \omega_p/v_F$.

\textit{Landau damping}. Obviously, the second term in Eq.~(\ref{3.15}) generally contains an imaginary part even in the collision\textit{less} limit where $\tau^{-1}$ is vanishingly small, due to a pole at $\bar{\omega} = \mathbf{K}\cdot\mathbf{v}$ in the integrand of the integral in $\mathbf{F}$. This part gives rise to Landau damping. For $\gamma_0/\omega_p\ll 1$, we find for $\bar{\omega} = \omega_s + i\gamma_0$ 
\begin{eqnarray}
&~&\Omega^2(K,\omega) \approx \Omega^2_0 + \frac{3\omega^2_p}{2} \mathcal{P}\int^1_{-1} dr \frac{r^3}{\omega_s/Kv_F - r} \nonumber \\ &~& \quad \quad \quad \quad \quad \quad \quad \quad - i \frac{3\pi \omega^2_p}{2}\left(\frac{\omega_s}{Kv_F}\right)^3\Theta(Kv_F-\omega_s). \label{ld}
\end{eqnarray}
Here $\mathcal{P}$ takes the principal value. The sign of the second line depends on the sign of $\gamma_0$. Only for $\gamma_0>0$, it is negative leading to damping, in line with causality. Equation (\ref{ld}) shows that, for bulk waves Landau damping exists only for sufficiently large $K$. For SPWs, however, Landau damping always exists, because $q$ runs over all positive values in the secular equation (\ref{2.11}). 

A major improvement of the SCM over the SRM comes through the quantity $G(\mathbf{K},\omega)$. In the SRM and its descendents, $G = G_0$ contains no symmetry breaking effects. In the SCM, one finds $G = G_0 + G_s$, where $G_s$ stems from $J_s(0)$ and is given by
\begin{eqnarray}
    &~& G_s(\mathbf{K},\omega) = 4i\bar{\omega}\left(\frac{m}{2\pi\hbar}\right)^3\int_>d^3\mathbf{v}~v_z\left(-e^2f'_0\right)\times \\
    &~& \quad \quad \quad \left[\beta F_0(\mathbf{k},\bar{\omega},\mathbf{v})-p\beta F_0(\mathbf{k},\bar{\omega},\mathbf{v}_-) + 2(p-1)F_+(\mathbf{K},\bar{\omega},\mathbf{v})\right], \nonumber
\end{eqnarray}
where $'>'$ indicates that the integral is restricted to departing electrons, i.e. $v_z\geq0$. Note that $G_s$ depends linearly on $p$. 

\begin{figure*}
\begin{center}
\includegraphics[width=0.95\textwidth]{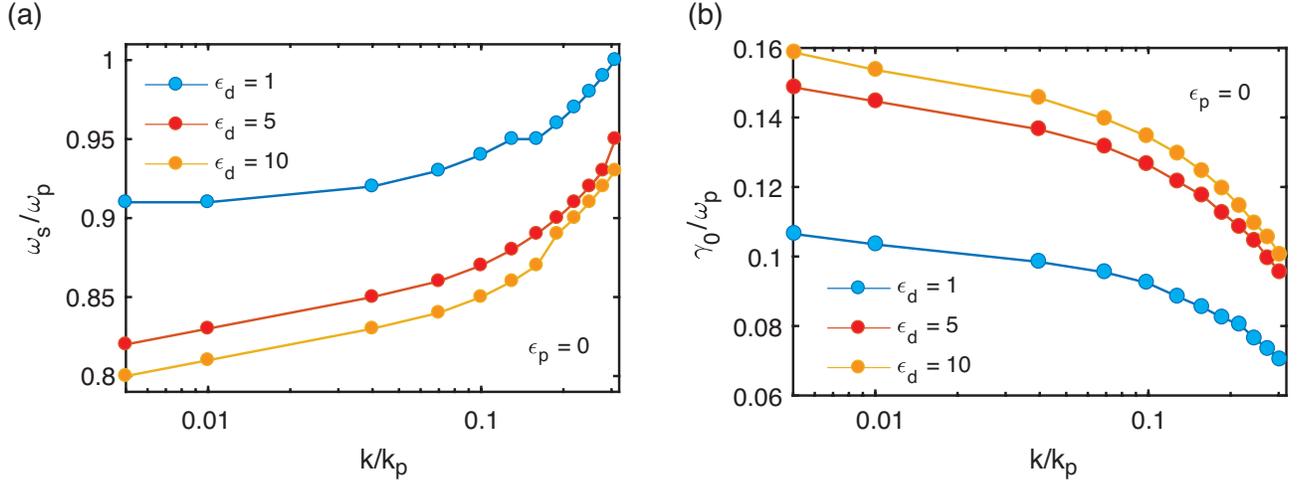}
\end{center}
\caption{Frequency $\omega_s$ and damping rate $\gamma = \tau^{-1}-\gamma_0$ of SPWs by the SCM. The results are obtained by numerically solving Eq.~(\ref{2.11}) without inter-band transition effects. $\omega_p$ is the characteristic frequency of the metal and $k_p = \omega_p/v_F$ with $v_F$ being the Fermi velocity. Solid lines are guides to the eye.\label{fig:f2}}
\end{figure*}  

\begin{figure}
\begin{center}
\includegraphics[width=0.45\textwidth]{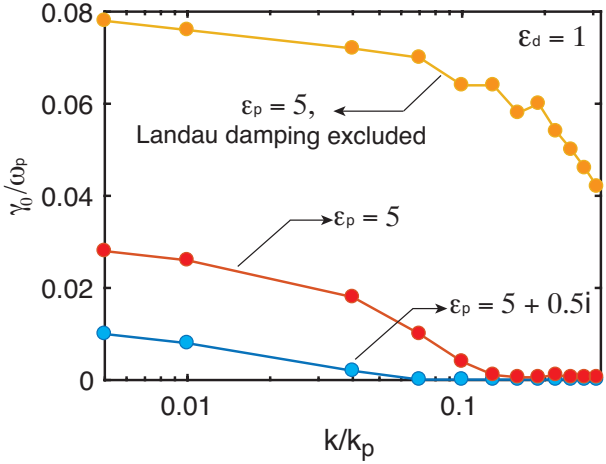}
\end{center}
\caption{Effects of inter-band absorption and Landau damping on $\gamma_0$. The results are obtained by numerically solving Eq.~(\ref{2.11}) for $\epsilon_d = 1$. Exclusion of Landau damping is done by dropping the imaginary part of $\Omega(K,\omega)$. Solid lines are guides to the eye.\label{fig:f3}}
\end{figure}

\subsection{The frequency and losses of SPWs}
\label{sec:4.3}
With $\Omega$ and $G$, we now proceed to solve Eq.~(\ref{2.11}) to find the frequency $\omega_s$ and damping rate $\gamma$ of SPWs in the SCM. 

\textit{Analytical analysis}. As said before, the most significant improvement of the SCM over the SRM is through the term $G_s$. We wish to do an approximate analysis to explicitly demonstrate how $G_s = G'_s + iG''_s$ affects $\omega_s$ and $\gamma$. 

For this purpose, let us write $G = G'+iG''$ and $\Omega = \Omega' + i\Omega''$ as well as $\epsilon = \epsilon'+i\epsilon''$, and assume that $\gamma_0/\omega_s$, $G''/G'$ and $\Omega''/\Omega'$ are all small quantities. Then, one can show that the real part of Eq.~(\ref{2.11}) gives
\begin{equation}
1+\frac{1}{\omega^2_s}\int^\infty_0 \frac{dq}{K^2} \frac{G'(\mathbf{K},\omega_s)}{\epsilon'(\mathbf{K},\omega_s)} \approx 0, \label{ws}
\end{equation}
which determines the SPW frequency $\omega_s$. The imaginary part of Eq.~(\ref{2.11}) yields $\gamma_0$ as
\begin{equation}
\gamma_0 \approx \omega_s\left(\eta_s - \eta_0\right),
\end{equation}
where the contribution 
\begin{equation}
\eta_s = \frac{1}{2}\int^\infty_0\frac{dq}{K^2}\frac{G''(\mathbf{K},\omega_s)}{\epsilon'(K,\omega_s)}/\int^\infty_0\frac{dq}{K^2}\frac{G'(\mathbf{K},\omega_s)}{\epsilon^{'2}(K,\omega_s)} \label{etas}
\end{equation}
stems directly from the imaginary part $G''$ -- which equals $G''_s$ -- of $G$, and
\begin{equation}
\eta_0 = \frac{\int^\infty_0\frac{dq}{K^2}\frac{G'(\mathbf{K},\omega_s)}{\epsilon'(K,\omega_s)}\frac{\Omega''(K,\omega)}{\Omega'(K,\omega)}\frac{\epsilon'(K,\omega_s)-1}{\epsilon'(K,\omega_s)}}{\int^\infty_0\frac{dq}{K^2}\frac{G'(\mathbf{K},\omega_s)}{\epsilon^{'2}(K,\omega_s)}} \label{eta0}
\end{equation}
comes directly from the imaginary part $\Omega''$ of $\Omega$. For stable systems, $\eta_0$ must be non-negative. 

Note that $\Omega''$ signifies Landau damping and inter-band absorption, as is clear from Eq.~(\ref{ld}). As such, we may further split $\eta_0 = \eta_\text{Landau} + \eta_{\text{interband}}$, so that the SPW damping rate becomes
\begin{equation}
\gamma \approx \tau^{-1} + \gamma_\text{Landau} + \gamma_\text{interband} - \gamma_s,
\end{equation}
where $\gamma_{\text{Landau},\text{interband},s} = \omega_s\eta_{\text{Landau},\text{interband},s}$, the first term represents Joule heating, the second and the third stand for Landau damping and inter-band absorption, respectively, while the last one is completely new due to $G'' = G''_s$. For models where $G_s = 0$, e.g. the DM, the HDM and the SRM, this new term is absent. 

For the SCM, however, $G_s$ is finite. Retaining only the first term in the series of $F_0$, we find
\begin{eqnarray}
G_s &\approx& - \frac{(1+p)\beta k \omega^2_p}{2\pi} \nonumber \\ &+& 8i\bar{\omega}(p-1)\left(\frac{m}{2\pi\hbar}\right)^3\int_>d^3\mathbf{v}~v_z\left(-e^2f'_0\right)F_+(\mathbf{K},\bar{\omega},\mathbf{v}). \nonumber
\end{eqnarray}
The second line here approximately corresponds to $iG''_s$. In general $G''_s\leq0$ and hence $\eta_s\geq0$. This implies that symmetry breaking effects tend to counteract the conventional damping and destabilize the metal. Our numerical solutions shall demonstrate that $\gamma_0$ is non-negative, in accord with the general argument given in Sec.~\ref{sec:4.1}.

In the long wavelength limit $k\sim 0$, one has $k/K^2 \approx \pi \delta(q)$. Under this approximation, Eq.~(\ref{ws}) becomes
\begin{equation}
\epsilon'(K_0,\omega_s) + \frac{\pi G'(\mathbf{K}_0,\omega_s)}{2k\omega^2_s} \approx 0. 
\end{equation}
Here $\mathbf{K}_0 = (\mathbf{k},0)$. For models with $G_s=0$, one immediately recovers from this equation the relation that $\epsilon_d + \epsilon' = 0$ as expected. For the SCM, however, one finds instead
\begin{equation}
\epsilon' + \epsilon_d\left(1-\frac{1+p}{2}\frac{\omega^2_p}{\omega^2_s}\right) = 0. 
\end{equation}
This leads to
\begin{equation}
\omega_s \approx \omega_p\left(\frac{1+\epsilon_d(1+p)/2}{1+\epsilon_{pr}+\epsilon_d}\right)^{1/2}. \label{3.19}
\end{equation}
This result differs considerably from what is expected of other models. It shows that the value of the SPW frequency depends on surface conditions.  

If we replace in Eq.~(\ref{etas}) $\epsilon'(K,\omega_s)$ with its non-dispersive part, as is reasonable for small $k$, then we find  
\begin{equation}
\eta_s\approx - \frac{1}{2\omega^2_{p}}\frac{1+\epsilon_d}{1+\epsilon_d(1+p)/2}\int^{\infty}_0 dq~\frac{G''_s(\mathbf{K},\omega_{s})}{K^2}, \label{3.18}
\end{equation}
which implies that inter-band transitions have little effect on $\eta_s$, whereas a dielectric can enhance it by as much as $200\%$. This is to be borne out in numerical analysis in what follows.

\textit{Numerical analysis}. We solve Eq.~(\ref{2.11}) numerically to find $\omega_s$ and $\gamma_0$ and how they vary with $k$ and $\epsilon_d$. We present the results for diffuse surfaces with $p=0$ only. In Eq.~(\ref{2.11}), the integral over $q$ extends to infinity. In numerical calculations, we impose a cutoff $q_c$. Roughly,  $q_c\sim a^{-1}$, where $a$ is a lattice constant. For metals, this means $q_c \sim k_F \sim k_p$. In all the numerical results presented here, we have chosen $q_c = 1.5k_p$. For $q_c$ beyond this value, both $\omega_s$ and $\gamma_0$ quickly converge, confirming that the results are independent of the choice of $q_c$~\cite{qc}. Our results should be taken with a grain of salt for very small $k\ll \omega_s/c \approx k_pv_F/c \sim 0.01k_p$, where $c$ is the speed of light in vacuum, because of retardation effects neglected in our theory. 

The results are displayed in Figs.~\ref{fig:f2} and \ref{fig:f3}. In Fig.~\ref{fig:f2}, we show $\bar{\omega} = \omega_s+i\gamma_0$ as a function of $k$ for various $\epsilon_d$ but without inter-band transition effects ($\epsilon_p=0$). As seen in Fig.~\ref{fig:f2} (a), in agreement with the analytical expression of $\omega_{s}$, increasing $\epsilon_d$ leads to smaller $\omega_s$. Note that $\omega_s$ is considerably larger than what would be obtained by other models due to surface effects. Meanwhile, $\gamma_0$ increases with increasing $\epsilon_d$, as seen in Fig.~\ref{fig:f2} (b), in accord with Eq.~(\ref{3.18}). This increase comes from the factor $\beta$, i.e. the presence of a dielectric enhances the electric field at the surface; see Eqs.~(\ref{2.15}) and (\ref{2.16}). 

The effects of inter-band absorption and Landau damping are illustrated in Fig.~\ref{fig:f3}. Here we plot $\gamma_0$ for $\epsilon_d=1$ under several circumstances as described in the figure. We see that inter-band transitions can strongly diminish $\gamma_0$ in two ways, as can be deduced from Eqs.~(\ref{3.19}) and (\ref{3.18}). Firstly, there is the screening effect (the curve with $\epsilon_p=5$). This leads to smaller $\omega_{s}$ and hence smaller $\gamma_0$, while leaving $\eta_0$ unaffected. Secondly, inter-band absorption further reduces $\gamma_0$ (see the curve with $\epsilon_p=5+0.5i$). As for Landau damping, it is sizable and generally increases with $k$; see Refs.~\cite{khurgin,deng2017c}. As such, $\gamma_0$ decreases as $k$ increases. In relation to this feature, we should mention a size effect~\cite{deng2017b}: in films of thickness $d$, $\gamma_0$ is strongly suppressed and quickly diminishes to zero when the wavelength becomes longer than $d$. Echoing this, one can show that $\gamma_0 \sim kv_F$ vanishes for $k\sim0$ in the SIM~\cite{deng2018}. 

\begin{figure*}
\begin{center}
\includegraphics[width=0.97\textwidth]{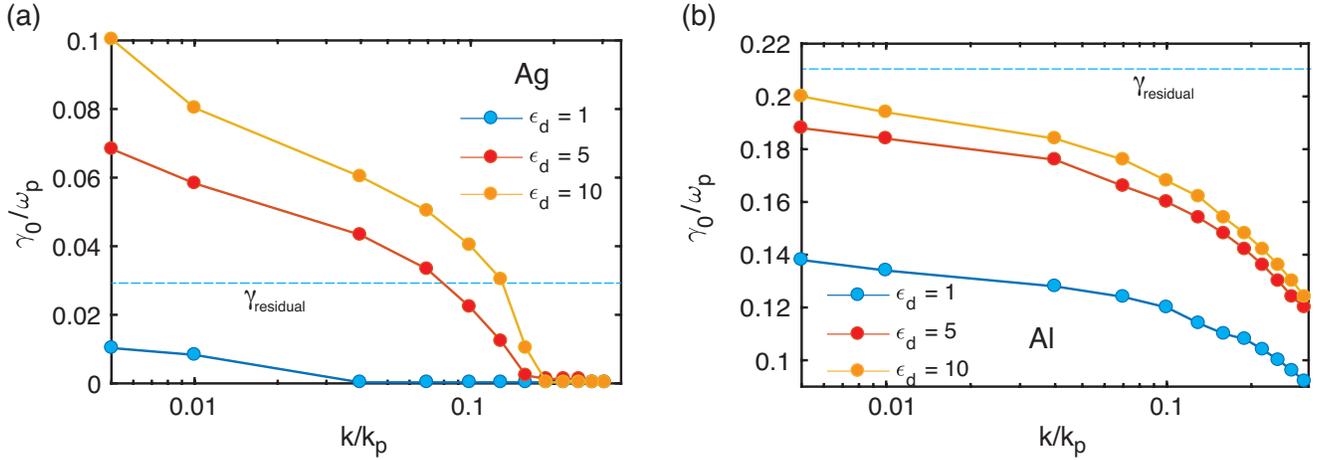}
\end{center}
\caption{$\gamma_0$ for SPWs in Ag and Al at various $\epsilon_d$. In both metals, $\gamma_0$ is enhanced with a dielectric. Solid lines are guides to the eye. \label{fig:f4}}
\end{figure*} 

\section{Possible Lossless SPWs by the SCM}
\label{sec:a3}
According to the SCM, the SPW loss rate is $\gamma = \tau^{-1}-\gamma_0$. In view of energy conversion, the expression implies a competition between the loss due to thermalization and the gain due to energy transferred from the electrons to the waves~\cite{deng2017c}. Should the condition $\gamma_0\tau = 1$ be fulfilled, lossless SPWs may be produced. In this section, we discuss this possibility against two common plasmonic metals: silver and aluminum. 

Note that $\tau^{-1}$ is the collision rate at the SPW frequency $\omega_s$. Even at zero temperature and in defect-free samples, there is sufficient phase space -- available due to the effective temperature $\hbar\omega_s/k_B$ -- for electronic scattering and thus $\tau^{-1}$ is comparable to $\omega_p$. Up to our knowledge, there is virtually no direct data on $\tau^{-1}$ for any materials. We then opt to estimate it by the following formula~\cite{beach1977}, 
\begin{equation}
\tau^{-1} \approx \gamma_\text{residual} + 4\nu_0\left(\frac{T}{T_D}\right)^5\int^{T_D/T}_0\frac{y^4dy}{e^y-1} + \omega_p\left(\frac{k_BT}{\hbar\omega_p}\right)^2, \label{37}
\end{equation}
where $\gamma_\text{residual}$ is the residual rate given by
\begin{equation}
\gamma_\text{residual} \approx \frac{2\nu_0}{5} + \frac{\omega_p}{4\pi^2}\left(\frac{\omega_s}{\omega_p}\right)^2 + \gamma_\text{impurity}. \label{36}
\end{equation}
Here the first contribution comes from phonon scattering, the second from electron-electron scattering and the third due to impurity scattering to be neglected hereafter. The second term generally underestimates the electron-electron rate in noble metals and Al by a few times~\cite{lawrence1976}. The coefficient $\nu_0\sim k_BT_D/\hbar$ may be determined from the slope of the phononic part of the D.C. conductivity at temperatures higher than the Debye temperature $T_D$. Equation (\ref{36}) shows that $\gamma_\text{residual}$ may well amount to quite a few percentages of $\omega_p$ for metals. 

\textit{Silver}. In silver electronic transitions involving the \textit{5s} and \textit{4d} bands have a dramatic effect on the properties of SPWs~\cite{liebsch1993,johnson1972,marini2002}, leading to $\hbar\omega_{s}\approx 3.69~$eV and $\hbar\omega_b \approx 3.92~$eV at long wavelengths, both lying far below the characteristic frequency $\hbar\omega_p = 9.48~$eV. Experimentally~\cite{beach1977,note4}, it was found that $\hbar\nu_0\sim 0.1~\text{eV}\sim 0.01\hbar\omega_p$, in consistency with the value of $T_D \approx 220~$K for Ag. The electron-electron scattering rate according to Eq.~(\ref{36}) would be less than one percent of $\omega_p$, while experimental measurements and more accurate expressions~\cite{lawrence1976} place it about $0.02\omega_p$. As such, we may reasonably take $\gamma_\text{residual}\sim0.03\omega_p$ as an estimate. 

The damping rate $\gamma$ can be read out from the line shape of the electron energy loss spectra (EELS). The temperature dependence of $\gamma$ has been recorded on a high-quality Ag single crystal by means of EELS~\cite{rocca1990,rocca1992}. The data indicates that at low temperature $\gamma$ amounts to less than one percent of $\omega_p$, a value a few times smaller than the as-projected $\gamma_\text{residual}$. If Landau damping and inter-band absorption are also counted, the discrepancy can be more dramatic. In light of the present theory, this discrepancy gives an estimate of $\gamma_0$. The values suggest that $\gamma_0$ has substantially compensated for the collision losses, i.e. $\gamma_0 \sim \gamma_\text{residual}$, as borne out in the following analysis. 

To evaluate $\gamma_0$, $\epsilon_p(\omega)$ needs to be supplied. Combining a semi-quantum model and ellipsometry as well as transmittance-reflectance measurements, Raki\'{c} \textit{et al.}~\cite{rakic1998} employed K-K analysis and prescribed a parametrized dielectric function -- $\hat{\epsilon}^{(b)}_r(\omega)$ in their notation -- for the inter-band contribution. Here, we use their fitting as an input for $\epsilon_p(\omega)$ but with a number of caveats. Firstly, their dielectric function was deduced from measurements assuming the conventional electromagnetic responses without any surface effects considered in the present paper. These effects, however, should be considered when analyzing ellipsometry and reflectance spectra. A future study will be made to address this issue. Secondly, their function very poorly reproduces the electron energy loss spectra and the reflectance spectra, especially near the SPW frequency of interest. Thirdly, their function does not give an accurate partition into inter-band and intra-band contributions, e.g. $\omega_p\approx 8$eV was used rather than the widely agreed $9.48$eV~\cite{marini2002}, which may overestimate the inter-band transition effects. Finally, their function is defined only for real frequency, and thus in general not suitable for $\epsilon_p(\omega)$, where $\omega$ is complex. To remedy this point, we substitute $\hat{\epsilon}^{(b)}_r[\text{Re}(\omega)]$ for $\epsilon_p(\omega)$, which should be reasonable if $\gamma/\omega_s\ll 1$. 

\begin{figure}
\begin{center}
\includegraphics[width=0.47\textwidth]{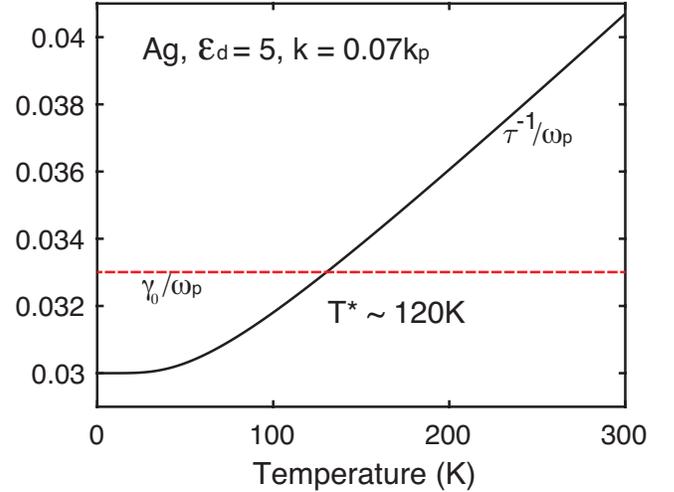}
\end{center}
\caption{The possibility of compensating for the SPW losses in Ag. A critical temperature $T^*$ exists where $\gamma_0 \tau=1$. Cooling down the system toward $T^*$ reduces the losses to a vanishingly small level. Here $\tau^{-1}$ is calculated by Eq.~(\ref{37}) with $T_D=220$K and $\hbar\nu_0 = 0.1$eV. \label{fig:f5}}
\end{figure} 

The results are displayed in Fig.~\ref{fig:f4} (a), where the computed $\gamma_0$ is exhibited as a function of $k$ at various values of $\epsilon_d$. While $\gamma_0$ for SPWs supported on a pristine Ag surface (i.e. $\epsilon_d = 1$) is negligibly small and way below $\gamma_\text{residual}$, by using a dielectric it can be significantly enhanced at long wavelengths beyond $\gamma_\text{residual}$. This trend is consistent with Eq.~(\ref{3.18}). The fact that $\gamma_0$ can be made higher than $\gamma_\text{residual}$ suggests the possibility of compensating for the plasmonic losses completely in Ag. The situation is shown in Fig.~\ref{fig:f5}. By interfacing the metal with a lossless dielectric of $\epsilon_d=5$ and cooling it down toward a critical temperature $T^*\sim 120$K, one can diminish the net losses as much as desired. It should be mentioned that, $\epsilon_d$ is the constant at the frequency $\omega_s$ as well. 

\textit{Aluminum}. Inter-band transitions in Al are widely considered less pronounced than in Ag. Nevertheless, their presence can still be felt, e.g. in the difference between the values of $\hbar\omega_p \approx 12.6$eV and $\hbar\omega_b \approx 15.3$eV. These numbers were obtained by density functional theory~\cite{lee1994} and experimental fitting~\cite{rakic1995}. In addition, $\hbar\omega_s\approx 10.7$eV~\cite{powell1959,knight2014}. $\gamma_\text{residual}$ may be deduced from the experimental measurements performed by Sinvani \textit{et al.}~\cite{sinvani1981} and others~\cite{ribot1979}. These authors measured the low temperature dependence of the d.c. resistivity $\rho$ of Al. Their data shows that $\rho \approx \rho_0 + AT^2 +B(T/T_D)^5$, where $\rho_0$ stems from impurity and lattice dislocation scattering while $A$ and $B$ are constants characterizing electron-electron scattering and electron-phonon scattering, respectively. Analyzing the data, the authors found that $A\approx 0.21$p$\Omega\cdot$cm$/$K$^2$ (a lower bound) and $B\approx 4.9\cdot 10^4\mu\Omega\cdot$cm for $T_D=430$K. From this we obtain $\nu_0\approx0.18\omega_p$ and the residual electron-electron scattering rate -- a few times larger than what would be obtained with Eq.~(\ref{36})~\cite{beach1977} --  approximating $0.14\omega_p$, yielding $\gamma_\text{residual}\approx 0.21\omega_p$. It is noted that this value is comparable to the width ($\sim 1.5$eV, nearly $0.12\omega_p$) of the electron energy loss peak near the frequency of the bulk plasma waves in Al~\cite{powell1959}. The as-obtained $\nu_0$ (and hence $\gamma_\text{residual}$) represents probably an overestimate~\cite{note5}. 

To compute $\gamma_0$, we again resort to the fitting function constructed by Raki\'{c} \textit{et al.}~\cite{rakic1995} and have it in place of $\epsilon_p(\omega)$ in our theory, in the same way as we did in the case of Ag. It goes without saying that the same caveats should be kept in mind. The results are shown in Fig.~\ref{fig:f4} (b). As expected, $\gamma_0$ is comparable to that in the absence of inter-band transitions [see Fig.~\ref{fig:f2} (b)], as these transitions are weak in Al. As is with Ag, $\gamma_0$ of SPWs in Al can also be fortified -- but to a lesser extent -- by a dielectric. Nevertheless, the enhanced $\gamma_0$ still falls short of $\gamma_\text{residual}$ unless for very long wavelengths where retardation effects need to be properly accounted for. 

The calculations reported in the above have assumed $p=0$. For surfaces strongly reflecting electrons (i.e. $p>0$), $\gamma_0$ could be much lower~\cite{deng2017c}. We also point out that additional losses such as due to SPWs converted into radiation are not considered. They can be absorbed in the definition of $\tau^{-1}$. 

\section{Conclusions}
\label{sec:5}
In order to answer the two questions posed at the beginning of this paper, i.e. (i) why had not previous work hit at the possibility of lossless SPWs and (ii) how far is the latter from reality, we have derived a universal macroscopic theory of SPWs that applies to any electron dynamics. In light of the theory, our answer to question (i) is simple: lossless waves are possible only within a self-consistent description of physical surfaces that takes care of translation symmetry breaking effects, a condition not met in existing work. As for question (ii), we can only suggest an optimistic prospect rather than an answer due to various uncertainties in inter-band transition effects: our estimate shows that lossless waves may well be within the reach in some materials.  

Our results reveal two contradictory views regarding SPW losses, as compared in Table~\ref{t:1}. According to the conventional framework, as exemplified by the DM, the HDM and the SRM, $$\gamma^{(1)} = \tau^{-1} + \gamma_\text{Landau} + \gamma_\text{interband}$$ and thus the SPW loss rate cannot be smaller than either of $\tau^{-1}$ and $\gamma_\text{Landau}$. On the other hand, within the SCM, a totally different picture emerges, giving $$\gamma^{(2)} = \tau^{-1} - \gamma_0, \quad \gamma_0 = \gamma_s - \gamma_\text{Landau} - \gamma_\text{interband},$$ which suggests that the loss rate is always smaller than $\tau^{-1}$. Here $\gamma_s = \eta_0\omega_s$, see Eq.~(\ref{3.19}). To directly contrast these two views, one has to measure separately $\tau^{-1}$ and $\gamma$. While the latter can be measured in many ways, the former is difficult to be directly measured. In what follows, we mention some indirect observations defying $\gamma^{(1)}$ but supporting $\gamma^{(2)}$. 

Firstly, we note that $\gamma^{(2)}$ naturally resolves a long-standing puzzle, that of the apparent insignificance of Landau damping even at very short wavelengths~\cite{raether1988}. For example, in single crystal Ag, the loss rate measured by EELS~\cite{rocca1990,rocca1992} is $\sim 1\%\omega_p$ even for $k \sim 1\text{nm}^{-1}$, whereas $\gamma_\text{Landau} \sim kv_F \sim 10\%\omega_p$. This discrepancy is inexplicable by $\gamma^{(1)}$, but easily comprehensible by $\gamma^{(2)}$, i.e. Landau damping has been overcompensated by $\gamma_s$. 

Secondly, we note that the loss rate of bulk plasma waves differs from that of SPWs primarily because of the absence of $\gamma_0$ in the former, as least in materials where inter-band effects are not important. In those materials, bulk waves should be generally much more lossy than SPWs, an observation that seems in consistency with experience. For example~\cite{beck}, the loss rates for the SPWs and the bulk waves in potassium are $0.1\text{eV}\hbar^{-1}$ and $0.24\text{eV}\hbar^{-1}$, respectively, while those in cesium are $0.23\text{eV}\hbar^{-1}$ and $0.75\text{eV}\hbar^{-1}$, respectively. In spite of these, the general situation is obviously unclear at this stage.

Finally, we mention an experiment performed on a van der Waals structure by Iranzo \textit{et al.}~\cite{iranzo2018}. These authors were able to confine propagating plasmon between a graphene layer and a metal array to the atomic limit without sacrificing its lifetime, which obviously beats the limit set by Landau damping. From an energy conversion point of view~\cite{deng2017c}, the plasmon in such a structure is not much different from the surface plasmon on a metal surface. Their result is compatible with $\gamma^{(2)}$: in ultimate confinement $\gamma_0$ tends to zero due to increase of Landau damping (Fig.~\ref{fig:f3}), leaving the loss rate saturating at $\tau^{-1}$, as observed. We anticipate a similar trend for the losses of local plasmon resonance on metal particles. 

In the SCM we have assumed that the ground state of the underlying metal be simply the Fermi sea. The fact that $\gamma$ can be made negative means an instability of the Fermi sea. Upon entering such circumstances, the metals are expected to undergo a transition into a different stable state, of which the electrical responses cannot be captured by our current SCM calculations. We will clarify the nature of this transition in the future.  

The results reported in this work should be of broad interest to the researchers working in plasmonics, surface science and condensed matter physics. We hope that experimentalists will find the results fascinating enough to put their hands on them. 

\

\textbf{Acknowledgement} -- The author enjoyed the hospitality during his stay with K. Wakabayashi's group at Kwansei Gakuin University, Japan, where part of the writing was undertaken. This work is not supported by any funding bodies. 

\appendix
\section{SPWs in the HDM}
\label{sec:a1}
Here we show that the usually claimed SPWs in the HDM are incompatible with the waves in the DM. We put $\epsilon_p = 0$ and $\epsilon_d = 1$ for simplicity. In the HDM, the electrons are treated as a fluid described by two field quantities: the velocity field $\mathbf{v}(\mathbf{x},t)$ and the electron density field $n_0 + n(\mathbf{x},t)$, where $n(\mathbf{x},t)$ denotes the deviation from the mean density $n_0$. The charge density is then $\rho(\mathbf{x},t) = n(\mathbf{x},t)e$ and the current density is $\mathbf{J}(\mathbf{x},t) = e(n_0+n(\mathbf{x},t))\mathbf{v}(\mathbf{x},t)$, which in the linear responses regime becomes $\mathbf{J}(\mathbf{x},t) = n_0e\mathbf{v}(\mathbf{x},t)$. 

\begin{figure*}
\begin{center}
\includegraphics[width=0.97\textwidth]{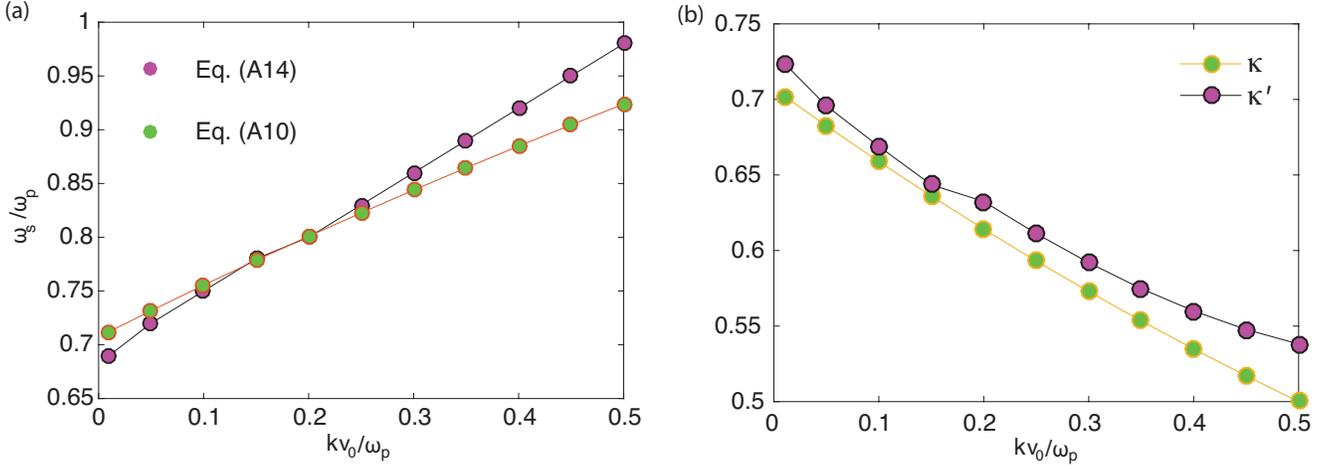}
\end{center}
\caption{SPWs in the HDM. In panel (a), the dispersion relation obtained by Eq.~(\ref{a14}) is plotted. The erroneous solution, Eq.~(\ref{a9}), which is widely quoted as the SPW dispersion, is shown alongside for comparison. As for the charge density, one has $\rho(z) \sim v^{-1}_0 e^{-\kappa' z}$ for $z$ not so close to the surface. The exponent $\kappa' \sim v^{-1}_0$ is shown in panel (b). The erroneous solution states that $\rho(z) \sim e^{-\kappa z}$, where $\kappa$ is shown also in (b).\label{fig:f6}}
\end{figure*} 

A small fluid element of volume $\delta V$ feels a force consisting of two portions: the electric force $n_0e\delta V \mathbf{E}(\mathbf{x},t)$ and the pressure due to density variation $-m\delta V v^2_0\partial_\mathbf{x}n(\mathbf{x},t)$. Now the laws of mechanics states that in the linear regime one has
\begin{equation}
    n_0m \left(\partial_t+\frac{1}{\tau}\right)\mathbf{v}(\mathbf{x},t) = n_0e\mathbf{E}(\mathbf{x},t) - mv^2_0\partial_\mathbf{x}n(\mathbf{x},t), 
\end{equation}
Here shear viscosity effects have been ignored. Now assuming $n(\mathbf{x},t) = n(\mathbf{x})e^{-i\omega t}$ and similarly for $\mathbf{v}(\mathbf{x},t)$ and other field quantities, we obtain the current density as 
\begin{equation}
    \mathbf{J}(\mathbf{x}) = \frac{i}{\bar{\omega}}\left(\frac{\omega^2_p}{4\pi}\mathbf{E}(\mathbf{x}) - v^2_0\partial_\mathbf{x}\rho(\mathbf{x})\right), \label{A2}
\end{equation}
the divergence of which is then given by
\begin{equation}
    \partial_\mathbf{x}\cdot\mathbf{J}(\mathbf{x}) = \frac{i}{\bar{\omega}}\left(\omega^2_p - v^2_0\partial^2\right)\rho(\mathbf{x}). \label{a3}
\end{equation}
Combining this equation with the equation of continuity, one obtains the wave equation for the charge density. 

\textit{The usually quoted SPWs}. In the standard but erroneous prescription for SPWs in this model, one takes $J_z(\mathbf{x}_0) \equiv 0$, or equivalently
\begin{equation}
    v_z(\mathbf{x}_0) \equiv 0. \label{a4}
\end{equation}
Here $\mathbf{x}_0$ denotes a point on the surface. The continuity equation reads
\begin{equation}
    -i\bar{\omega}\rho(\mathbf{x}) + \partial_\mathbf{x}\cdot\mathbf{J}(\mathbf{x}) = 0. \label{A5}
\end{equation}
In conjunction with Eq.~(\ref{a3}), one finds
\begin{equation}
    \left(\omega^2_p - \bar{\omega}^2 - v^2_0\partial^2\right)\rho(\mathbf{x}) = 0, \label{a6}
\end{equation}
which is Eq.~(\ref{2.20}) given in Sec.~\ref{sec:2.3}. One then seeks solutions of the form $\rho(\mathbf{x}) = \rho(z) e^{ikx}$ and similarly for other quantities. Further, taking $\rho(z) = \rho_0e^{-\kappa z}$ and substituting it in Eq.~(\ref{a6}), we obtain
\begin{equation}
    \omega^2_p + v^2_0(k^2-\kappa^2) - \bar{\omega}^2 = 0. 
\end{equation}
The boundary condition (\ref{a4}) requires
\begin{equation}
    \frac{\omega^2_p}{4\pi}E_z(0) = v^2_0 ~ \rho'(0) = -\kappa v^2_0 \rho_0, \quad \rho'(z) = \partial_z\rho(z). \label{a8}
\end{equation}
It is easy to show that $$ E_z(0) = -\frac{2\pi \rho_0}{k+\kappa}.$$ With this it follows from Eq.~(\ref{a8}) that
\begin{equation}
    \omega^2_0 = v^2_0\kappa (k+\kappa), \quad \omega_0 = \omega_p/\sqrt{2}. 
\end{equation}
Combining this relation with Eq.~(\ref{a6}), we arrive at
\begin{equation}
    \bar{\omega}^2 = \omega^2_0 + \omega_0 v_0 k. \label{a9}
\end{equation}
This is the usually quoted dispersion relation claimed for the SPWs in the HDM. As briefly captured in Sec.~\ref{sec:2.3}, this claim is plainly false: in the limit $v_0 \rightarrow 0$, $\kappa \sim v^{-1}_0$ diverges and hence $\int^\infty_0 dz ~ \rho(z) = \rho_0\kappa^{-1} \sim \rho_0 v_0$ vanishes, thus in contradiction with the DM. This would also erroneously imply that $E_z(z)$ did not change sign across the surface charge layer. It was essentially this erroneous solution that had been identified by Harris~\cite{harris1972} and Garcia \textit{et al}~\cite{garcia1977b} in their study based on Boltzmann's equation. It is plausible that this is also so with works employing a more microscopic approach such as the density functional theory, at least those using the so-called 'infinite barrier' model for mimicking the surface~\cite{pitarke2001}. For example, Feibelman identified a solution of uniform potential and hence empty of charges but with frequency $\omega_p/\sqrt{2}$ in the long wavelength limit~\cite{feibelman1971}, exactly in this kind.

\textit{SPWs in the HDM by the present theory}. In our theory, no restrictions are placed on $J_z(\mathbf{x}_0)$ and thus the equation of continuity reads
\begin{equation}
    -i\bar{\omega}\rho(\mathbf{x}) + \partial_\mathbf{x}\cdot\mathbf{J}(\mathbf{x}) = - \Theta'(z) J_z(\mathbf{x}_0). \label{a11}
\end{equation}
Using Eq.~(\ref{a3}) and taking $\rho(\mathbf{x}) = \rho(z) e^{ikx}$ and similarly for other quantities, one finds
\begin{equation}
    \left(\omega^2_p+v^2_0k^2 -v^2_0\partial^2_z - \bar{\omega}^2\right)\rho(z) = i\bar{\omega}J_z(0) \Theta'(z).
\end{equation}
Now we introduce the Fourier transform for $$\rho(z) = (2/\pi) \int^\infty_0 dq ~ \rho_q \cos(qz).$$ It follows that $$J_z(0) = (i/\bar{\omega})(\omega^2_p/4\pi) E_z(0) = \sigma_\text{DM}E_z(0).$$ This implies that $\rho'(0) = 0$, which must hold for any $\rho(z)$ that reduces to the Dirac function $\delta(z)$ desired in the DM limit. In conjunction with 
\begin{equation}
    E_z(0) = - \int^\infty_0 dq \frac{4 k \rho_q}{K^2}, 
\end{equation}
we find from Eq.~(\ref{a11}) that
\begin{equation}
    1 - \frac{\omega^2_p}{4\pi} \int^\infty_0 dq~\frac{k}{K^2}\frac{1}{\omega^2_p + v^2_0K^2 - \bar{\omega}^2} = 0, \label{a14}
\end{equation}
which is just Eq.~(\ref{2.181}) displayed in Sec.~\ref{sec:2.3}. The dispersion relation obtained by this equation has been plotted in Fig.~\ref{fig:f6} (a), where the relation~(\ref{a9}) is displayed together for comparison. Both exhibit a linear dependence on $k$ but with different slopes. Our theory predicts a slightly bigger slope. 

Let us have a look at the profile of $\rho(z)$. According to our theory, $$\rho(z) = i\bar{\omega}J_z(0)\int \frac{dq}{\pi/2} \cos(qz) \left(\omega^2_p + v^2_0K^2-\bar{\omega}^2\right)^{-1}.$$ With $\tilde{q} = q v_0/\omega_p$ and $\tilde{K} = Kv_0/\omega_p$ as well as $\tilde{z} = z\omega_p/v_0$, it can be rewritten as
\begin{equation}
    \rho(z) = \frac{i\bar{\omega}J_z(0)}{\omega_pv_0} \tilde{\rho}(z), 
\end{equation}
where 
\begin{equation}
\tilde{\rho}(z) = \int \frac{d\tilde{q}}{\pi/2} \cos(\tilde{q}\tilde{z}) \left(1+\tilde{K}^2 - (\bar{\omega}/\omega_p)^2\right)^{-1}. \label{a16}  
\end{equation}
One can show that not so close to the surface $\tilde{\rho}(z) = \frac{1}{\kappa'v_0/\omega_p}e^{-\kappa'z}$, where $$\kappa' = \frac{\omega_p}{v_0}\sqrt{1 + \tilde{k}^2 -(\bar{\omega}/\omega_p)^2}$$ with $\tilde{k} = kv_0/\omega_p$. Thus, $$\int \rho(z) ~ dz = \frac{i\bar{\omega}J_z(0)}{\omega^2_p}\frac{1}{1+\tilde{k}^2-(\bar{\omega}/\omega_p)^2}.$$ For $\tilde{k}=0$, it reduces to that for the DM. The dependence of $\kappa'$ on $k$ is shown in Fig.~\ref{fig:f6} (b), alongside that of $\kappa$.   

\section{Origin of SPWs in the SRM}
\label{sec:a2}
The widely used SRM assumes that the electrons be specularly reflected off a surface. One would then expect that $J_z=0$ and no SPWs would exist in this model. However, in the main text we have shown that what the SRM actually does is an extension of the HDM. The question then is, how do SPWs originate in the SRM? Here we show that the answer lies with the 'fictitious' charge sheet assumed in the model. 

We follow the SRM formalism as explained in many papers~\cite{ritchie1966,garcia1979,aminov2001,moliner,arista1992,yubero1996} and employ it to study the response to an external distribution of charge $\rho_\text{ext}(z;\mathbf{k},\omega)$ placed outside the metal. We take $\rho_\text{ext}(z;\mathbf{k},\omega) = \rho_0(\mathbf{k},\omega)\delta(z-z_0)$ for simplicity, where $z_0<0$. In the SRM, the total electrostatic potential, $\phi_\text{tot}(z;\mathbf{k},\omega) = \phi(z;\mathbf{k},\omega) + \phi_\text{ext}(z;\mathbf{k},\omega)$, where $\phi$ is the potential produced by the induced charges $\rho(z;\mathbf{k},\omega)$, is written
\begin{equation}
\phi_\text{tot}(z;\mathbf{k},\omega) = \Theta(z)\phi_m(z;\mathbf{k},\omega) + \Theta(-z)\phi_v(z;\mathbf{k},\omega), \label{f1}
\end{equation}
where $\phi_m$ and $\phi_v$ are the potentials in the so-called pseudo-metal and pseudo-vacuum, respectively. These are further defined by
\begin{equation}
\phi_{m/v}(\mathbf{Q},\omega) = \frac{4\pi}{Q^2\epsilon_{m/v}(\mathbf{Q},\omega)}\left[\rho^{m/v}_\text{ext}(\mathbf{Q},\omega)+\sigma^{m/v}_s(\mathbf{k},\omega)\right],
\end{equation}
where $\phi_{m/v}(\mathbf{Q},\omega) = \int^\infty_{-\infty} dz ~ e^{-iqz} \phi_{m/v}(z;\mathbf{k},\omega)$ is the ordinary Fourier transform, $\mathbf{Q} = (\mathbf{k},q)$, $\epsilon_m(\mathbf{Q},\omega) = \epsilon(Q,\omega)$, $\epsilon_v(\mathbf{Q},\omega) = 1$ and $\sigma^{m/v}_s(\mathbf{k},\omega)$ is the fictitious surface charge density. In addition, $\rho^{m/v}_\text{ext}$ is related to $\rho_\text{ext}$ as follows
\begin{equation}
\rho^{m/v}_\text{ext}(z;\mathbf{k},\omega) = \Theta(z)\rho_\text{ext}(\mp z;\mathbf{k},\omega) + \Theta(-z)\rho_\text{ext}(\pm z;\mathbf{k},\omega). \label{f3}
\end{equation}
It follows that $\rho^v_\text{ext}(\mathbf{Q},\omega) = 0$ and
\begin{equation}
\rho^m_\text{ext}(\mathbf{Q},\omega) = 2\rho_0(\mathbf{k},\omega)\cos(qz_0). 
\end{equation}
Equations (\ref{f1}) - (\ref{f3}) define the SRM. Requiring the continuity of the dielectric displacement at $z=0$ leads to 
\begin{equation}
\sigma^m_s(\mathbf{k},\omega) = -\sigma^v_s(\mathbf{k},\omega) = \sigma_s(\mathbf{k},\omega). 
\end{equation}
This can be further fixed by requiring the continuity of $\phi_\text{tot}$ at $z=0$. One finds
\begin{equation}
\sigma_s(\mathbf{k},\omega) = 2\rho_0(\mathbf{k},\omega)\cosh(kz_0)\epsilon^{-1}_\text{s,SRM}(\mathbf{k},\omega),
\end{equation}
where $\epsilon_\text{s,SRM}$ is as given in Sec.~\ref{sec:3.3}, i.e. $$\epsilon_\text{s,SRM}(\mathbf{k},\omega) = 1+\frac{k}{\pi}\int^\infty_{-\infty}\frac{dq}{Q^2\epsilon(Q,\omega)}.$$ The zeros of this quantity give poles of $\sigma_s$ corresponding to SPWs in the SRM, thus revealing that the fictitious charge sheet is responsible for the SPWs in the SRM.


\begin{thebibliography}{9}
\bibitem{qc}
The solutions to Eq.~(\ref{2.11}) are independent of the choice of $q_c$ as long as the latter is sufficiently large. For example, for $k = 0.07k_p$, we find that $\bar{\omega}/\omega_p = 0.93+0.095i, 0.95+0.099i, 0.95+0.099i, 0.96+0.098$ for $q_c/k_p = 1.5, 3.0, 4.5, 6.0$, respectively. There is no difference within the numerical resolution $\pm 0.005$. More discussions on this can be found in Ref.~\cite{deng2018}.
\end{thebibliography}
\end{document}